\documentclass[aps, prd, reprint,showpacs,  preprintnumbers,amsmath,amssymb,superscriptaddress, nofootinbib]{revtex4-1}
\usepackage{amssymb, amsmath, braket, nicefrac, graphicx} 
\usepackage{relsize}
\usepackage{lineno}
\usepackage{setspace}

\usepackage{ esint }
\usepackage{chngpage}
\usepackage{epsfig}
\usepackage{multirow}
\usepackage{rotating}
\usepackage{float}
\usepackage{color}
\usepackage{xcolor}
\usepackage{appendix}
\usepackage{ amstext }
\usepackage{ gensymb }
\usepackage{ textcomp }
\usepackage{ wasysym }
\usepackage{pstricks}

\usepackage[caption=false,labelformat=parens]{subfig}

\usepackage{dcolumn}
\usepackage{bm}
\usepackage{units}

\newcommand{\numu}{\mbox{$\nu_{\mu}$}}                   

\newcommand{\anumu}{\ensuremath{\bar{\nu}_{\mu}}}

\newcommand{\piz}{\mbox{$\pi^{0}$}} 
\newcommand{\simgt}{\,\hbox{\lower0.6ex\hbox{$\sim$}\llap{\raise0.6ex\hbox{$>$}}}\,}
\newcommand{\simlt}{\,\hbox{\lower0.6ex\hbox{$\sim$}\llap{\raise0.6ex\hbox{$<$}}}\,}

\newcommand{\minerva}{MINERvA~}

\definecolor{maroon}{RGB}{162,10,10}

\usepackage{hyperref}
\usepackage[all]{hypcap}
\hypersetup{
    colorlinks,
    citecolor=black,
    filecolor=black,
    linkcolor=black,
    urlcolor=black
}

\makeatletter

\renewenvironment{figure}
  {\def\@captype{figure}}
  {}
\makeatother

\begin{document}
\preprint{FERMILAB-PUB-17-220-ND}
\title{Measurement of $\numu$ charged-current single $\pi^{0}$ production on hydrocarbon in the few-GeV region using \minerva}

\newcommand{\Rutgers}{Rutgers, The State University of New Jersey, Piscataway, New Jersey 08854, USA}
\newcommand{\Hampton}{Hampton University, Dept. of Physics, Hampton, VA 23668, USA}
\newcommand{\Dortmund}{Institute of Physics, Dortmund University, 44221, Germany }
\newcommand{\Otterbein}{Department of Physics, Otterbein University, 1 South Grove Street, Westerville, OH, 43081 USA}
\newcommand{\JMU}{James Madison University, Harrisonburg, Virginia 22807, USA}
\newcommand{\Florida}{University of Florida, Department of Physics, Gainesville, FL 32611}
\newcommand{\UCIrvine}{Department of Physics and Astronomy, University of California, Irvine, Irvine, California 92697-4575, USA}
\newcommand{\CBPF}{Centro Brasileiro de Pesquisas F\'{i}sicas, Rua Dr. Xavier Sigaud 150, Urca, Rio de Janeiro, Rio de Janeiro, 22290-180, Brazil}
\newcommand{\PUCP}{Secci\'{o}n F\'{i}sica, Departamento de Ciencias, Pontificia Universidad Cat\'{o}lica del Per\'{u}, Apartado 1761, Lima, Per\'{u}}
\newcommand{\INRM}{Institute for Nuclear Research of the Russian Academy of Sciences, 117312 Moscow, Russia}
\newcommand{\Jlab}{Jefferson Lab, 12000 Jefferson Avenue, Newport News, VA 23606, USA}
\newcommand{\Pittsburgh}{Department of Physics and Astronomy, University of Pittsburgh, Pittsburgh, Pennsylvania 15260, USA}
\newcommand{\Guanajuato}{Campus Le\'{o}n y Campus Guanajuato, Universidad de Guanajuato, Lascurain de Retana No. 5, Colonia Centro, Guanajuato 36000, Guanajuato M\'{e}xico.}
\newcommand{\Athens}{Department of Physics, University of Athens, GR-15771 Athens, Greece}
\newcommand{\Tufts}{Physics Department, Tufts University, Medford, Massachusetts 02155, USA}
\newcommand{\WM}{Department of Physics, College of William \& Mary, Williamsburg, Virginia 23187, USA}
\newcommand{\FNAL}{Fermi National Accelerator Laboratory, Batavia, Illinois 60510, USA}
\newcommand{\Purdue}{Department of Chemistry and Physics, Purdue University Calumet, Hammond, Indiana 46323, USA}
\newcommand{\MCLA}{Massachusetts College of Liberal Arts, 375 Church Street, North Adams, MA 01247}
\newcommand{\UMD}{Department of Physics, University of Minnesota -- Duluth, Duluth, Minnesota 55812, USA}
\newcommand{\Northwestern}{Northwestern University, Evanston, Illinois 60208}
\newcommand{\UNI}{Universidad Nacional de Ingenier\'{i}a, Apartado 31139, Lima, Per\'{u}}
\newcommand{\Rochester}{University of Rochester, Rochester, New York 14627 USA}
\newcommand{\Austin}{Department of Physics, University of Texas, 1 University Station, Austin, Texas 78712, USA}
\newcommand{\USM}{Departamento de F\'{i}sica, Universidad T\'{e}cnica Federico Santa Mar\'{i}a, Avenida Espa\~{n}a 1680 Casilla 110-V, Valpara\'{i}so, Chile}
\newcommand{\Geneva}{University of Geneva, 1211 Geneva 4, Switzerland}
\newcommand{\Chicago}{Enrico Fermi Institute, University of Chicago, Chicago, IL 60637 USA}
\newcommand{\hired}{}
\newcommand{\OregonState}{Department of Physics, Oregon State University, Corvallis, Oregon 97331, USA}
\newcommand{\Wroclaw}{University of Wroclaw, plac Uniwersytecki 1, 50-137 Wroc?aw, Poland}
\newcommand{\oxford}{}
\newcommand{\bmeThanks}{now at SLAC National Accelerator Laboratory, Stanford, CA 94309, USA}
\newcommand{\higueraThanks}{now at University of Houston, Houston, TX 77204, USA}
\newcommand{\damartinezThanks}{now at Illinois Institute of Technology, Chicago, IL 60616, USA}
\newcommand{\mcgivernThanks}{now at Fermi National Accelerator Laboratory, Batavia, IL 60510, USA}
\newcommand{\joelmousseauThanks}{now at University of Michigan, Ann Arbor, MI 48109, USA}
\newcommand{\LazaThanks}{also at Department of Physics, University of Antananarivo, Madagascar}
\newcommand{\twaltonThanks}{now at Fermi National Accelerator Laboratory, Batavia, IL 60510, USA}
\newcommand{\jwolcottThanks}{now at Tufts University, Medford, MA 02155, USA}
\newcommand{\zavalaThanks}{Deceased}

\author{O.~Altinok}                       \affiliation{\Tufts}
\author{T.~Le}                            \affiliation{\Tufts}  
\author{L.~Aliaga}                        \affiliation{\WM}  \affiliation{\PUCP}
\author{L.~Bellantoni}                    \affiliation{\FNAL}
\author{A.~Bercellie}                     \affiliation{\Rochester}
\author{M.~Betancourt}                    \affiliation{\FNAL}
\author{A.~Bodek}                         \affiliation{\Rochester}
\author{A.~Bravar}                        \affiliation{\Geneva}
\author{H.~Budd}                          \affiliation{\Rochester}
\author{G.F.R.~Caceres~Vera}   \affiliation{\CBPF}
\author{T.~Cai}                           \affiliation{\Rochester}
\author{M.F.~Carneiro}                    \affiliation{\CBPF}
\author{H.~da~Motta}                      \affiliation{\CBPF}
\author{S.A.~Dytman}                      \affiliation{\Pittsburgh}
\author{G.A.~D\'{i}az~}                   \affiliation{\Rochester}  \affiliation{\PUCP}
\author{J.~Felix}                         \affiliation{\Guanajuato}
\author{L.~Fields}                        \affiliation{\FNAL}  \affiliation{\Northwestern}
\author{R.~Fine}                          \affiliation{\Rochester}
\author{A.M.~Gago}                 \affiliation{\PUCP}
\author{R.Galindo}                        \affiliation{\USM}
\author{H.~Gallagher}                     \affiliation{\Tufts}
\author{A.~Ghosh}                          \affiliation{\USM}
\author{R.~Gran}                          \affiliation{\UMD}
\author{J.Y.~Han}                      \affiliation{\Pittsburgh}
\author{D.A.~Harris}                      \affiliation{\FNAL}
\author{J.~Kleykamp}                      \affiliation{\Rochester}
\author{M.~Kordosky}                      \affiliation{\WM}
\author{E.~Maher}                         \affiliation{\MCLA}
\author{S.~Manly}                         \affiliation{\Rochester}
\author{W.A.~Mann}                        \affiliation{\Tufts}
\author{C.M.~Marshall}                    \affiliation{\Rochester}
\author{D.A.~Martinez~Caicedo}\thanks{\damartinezThanks}  \affiliation{\CBPF}
\author{K.S.~McFarland}                   \affiliation{\Rochester}  \affiliation{\FNAL}
\author{A.M.~McGowan}                     \affiliation{\Rochester}
\author{B.~Messerly}                      \affiliation{\Pittsburgh}
\author{J.~Miller}                        \affiliation{\USM}
\author{A.~Mislivec}                      \affiliation{\Rochester}
\author{J.G.~Morf\'{i}n}                  \affiliation{\FNAL}
\author{D.~Naples}                        \affiliation{\Pittsburgh}
\author{J.K.~Nelson}                      \affiliation{\WM}
\author{A.~Norrick}                       \affiliation{\WM}
\author{Nuruzzaman}                       \affiliation{\Rutgers}  
\author{V.~Paolone}                       \affiliation{\Pittsburgh}
\author{C.E.~Patrick}                     \affiliation{\Northwestern}
\author{G.N.~Perdue}                      \affiliation{\FNAL}  \affiliation{\Rochester}
\author{M.A.~Ramirez}                     \affiliation{\Guanajuato}
\author{R.D.~Ransome}                     \affiliation{\Rutgers}
\author{H.~Ray}                           \affiliation{\Florida}
\author{L.~Ren}                           \affiliation{\Pittsburgh}
\author{D.~Rimal}                         \affiliation{\Florida}
\author{P.A.~Rodrigues}                   \affiliation{\Rochester}
\author{D.~Ruterbories}                   \affiliation{\Rochester}
\author{H.~Schellman}                     \affiliation{\OregonState}  \affiliation{\Northwestern}
\author{J.T.~Sobczyk}                     \affiliation{\Wroclaw}
\author{C.J.~Solano~Salinas}              \affiliation{\UNI}
\author{M.~Sultana}                        \affiliation{\Rochester}
\author{S.~S\'{a}nchez~Falero}            \affiliation{\PUCP}
\author{E.~Valencia}                      \affiliation{\Guanajuato}
\author{J.~Wolcott}\thanks{\jwolcottThanks}  \affiliation{\Rochester}
\author{B.~Yaeggy}                         \affiliation{\USM}   

\date{\today}
\pacs{13.15.+g, 14.20.Gk, 14.60.Lm}

\begin{abstract}

The semi-exclusive channel $\nu_{\mu}+\textrm{CH}\rightarrow\mu^{-}\pi^{0}+\textrm{nucleon(s)}$
is analyzed using MINERvA exposed to the low-energy NuMI $\numu$ beam with spectral peak at $E_{\nu} \simeq 3$\,GeV.
Differential cross sections for muon momentum and production angle, $\pi^{0}$ kinetic energy and production angle,
and for squared four-momentum transfer are reported,  and the cross section $\sigma(E_{\nu})$ is obtained over the 
range 1.5\,GeV $\leq E_{\nu} <$ 20\,GeV.  Results are compared to GENIE and NuWro predictions
and to published MINERvA cross sections for charged current $\pi^+ (\pi^0)$ production by $\nu_{\mu} (\bar{\nu}_{\mu})$ neutrinos.
Disagreements between data and simulation are observed at very low and relatively high 
values for muon angle and for $Q^2$ that may reflect shortfalls in modeling of interactions on carbon.
For $\pi^{0}$ kinematic distributions however, the data are consistent with the simulation and 
provide support for generator treatments of pion intranuclear scattering.
Using signal-event subsamples that have reconstructed protons
as well as $\pi^{0}$ mesons, the $p\pi^{0}$ invariant mass distribution is obtained, 
and the decay polar and azimuthal angle distributions in the rest
frame of the $p\pi^{0}$ system are measured 
in the region of $\Delta(1232)^+$ production, $W < 1.4$\,GeV.
 \end{abstract}

\maketitle


\section{Introduction}

Production of single $\pi^{0,+}$ mesons by $\numu$ charged-current (CC) 
inelastic scattering on nuclei in the few-GeV region of 
neutrino energy, $E_{\nu}$, arises from three types of processes.   
For $E_{\nu}\,\leq 3$\,GeV, CC($\pi$) reactions are mostly
instances of neutrino-nucleon scattering wherein a bound nucleon 
is struck and caused to transition into a baryon resonance 
that promptly decays into a pion and a nucleon.
Production of the $\Delta$(1232) $P_{33}$ resonance is prominent 
and causes CC($\pi$) cross sections to rise rapidly from thresholds.  
As the incident $E_{\nu}$ is increased however, contributions from higher-mass $N^*$ resonances 
such as the $P_{11}$(1440), $D_{13}$(1520), and $S_{11}$(1535) states become significant.
Increasing $E_{\nu}$ also facilitates the onset of deep inelastic scattering, in which the neutrino
interacts with a valence or sea quark within a bound nucleon and the exiting quark hadronizes into one or multiple pions.   
Single pion production can also arise from neutrino-nucleon scattering that does not involve a resonance.  Such processes  
are referred to as nonresonant pion production and are often treated as a subsample of DIS processes that have final-state
hadronic invariant masses $W$ less than 1.7 GeV~\cite{Gallagher-2006}. 

Charged-current single pion production features prominently among neutrino interactions that occur within or around
the appearance peaks or the disappearance minima in events-vs-$E_{\nu}$ spectra measured by 
the long baseline neutrino oscillation experiments.     This is especially the case for NOvA and DUNE
~\cite{NOvA-expt, DUNE-expt} since the $\numu$ fluxes of these experiments have maxima 
at 2.0\,GeV and 2.5\,GeV respectively.   Improved knowledge of CC($\pi$) is also of keen interest
for the T2K and HyperK long baseline experiments~\cite{T2K-expt, HyperK-expt} 
whose flux spectra peak below 1.0 GeV, 
especially with regard to $\Delta$(1232) and nonresonant pion contributions. 
In long baseline oscillation experiments, neutrino event energies need to be measured to precisions of e.g.
$\leq 100$ MeV for DUNE and $\leq 50$ MeV for T2K~\cite{Theory_Mosel} in order to resolve
the neutrino-sector CP violating phase and of the ordering of the neutrino mass states.
Consequently new measurements of $\textrm{CC}(\pi)$ reactions on hydrocarbon and on other nuclei 
in the few-GeV region of $E_{\nu}$ are urgently needed.   
Such measurements enable the testing and refinement of the neutrino event generators, e.g. 
GENIE~\cite{Andreopoulos-NIM-2010}, NEUT~\cite{ref:NEUT}, NuWro~\cite{ref:NuWro}, and GIBUU~\cite{ref:GIBUU}. 
The predictions of these generators are a crucial element 
in physics evaluations of oscillated neutrino event spectra.

Among the CC($\pi$) reactions, the channel
\begin{equation}
\label{signal-channel}
\nu_{\mu}+\textrm{CH}\rightarrow\mu^{-}+\pi^{0}+X(\textrm{nucleons})
\end{equation}
is of particular interest.   In contrast to charged pion production,  
the events of channel \eqref{signal-channel} 
have electromagnetic showers as the dominant part 
of the visible hadronic system.  As a result, channel  \eqref{signal-channel}  provides useful information
about the responses of detector systems
to both signal and backgrounds for the electron-neutrino 
oscillation appearance measurements, $\nu_{\mu}\rightarrow\nu_{e}$.  

While the cross section for CC($\pi^{0}$) is smaller than for CC($\pi^+$) processes, the single-$\pi^{0}$ channel
is devoid of CC single pion coherent scattering (e.g. $\numu + C \rightarrow \mu^- \pi^+ + C$), 
consequently its final-state distributions provide an unencumbered
view of $W^+$ excitation of the nucleon~\cite{Mosel-PRC91-2015, Katori-2016}.    
Thus for example a four-momentum transfer, $Q^2$, distribution measured for CC($\pi^{0}$) is directly
representative of the channel, whereas a $Q^2$ distribution obtained with CC($\pi^+$) events will include an elevated
event rate at low $Q^2$ arising from coherent $\pi^+$ production -- a very different reaction type that may necessitate a further
accounting~\cite{Carrie-pion}.  An additional simplification is that single-$\pi^{0}$ production is composed of carbon-target scattering almost
entirely.   According to the reference simulation of this analysis, scatters from hydrogen account for less than $3.3\%$ of the candidate signal sample.
The kinematic spectra of the $\pi^{0}$ mesons of \eqref{signal-channel} provide useful checks on 
simulation treatments of pion final-state interactions (FSI) that can take place within the struck carbon nucleus.    The 
$\pi^{0}$ distributions in kinetic energy and production angle probe the suite of elastic, inelastic, absorption, and charge-exchange
scattering algorithms that are used in these treatments in ways that complement constraints 
gleaned from charged pion distributions~\cite{Brandon-pion, Trung-pion}.

\smallskip

In the present work, a signal event sample for channel \eqref{signal-channel} is isolated which, after background 
subtraction, is predicted by the reference model to have a 50\% contribution from the exclusive channel
\begin{equation}
\label{exclusive-channel}
\nu_{\mu} + \textrm{n} \rightarrow\mu^{-}+\pi^{0}+\textrm{p}
\end{equation}
where the struck neutron is bound within a carbon nucleus.
Thus the phenomenology pertaining to reaction \eqref{exclusive-channel} can be helpful when
evaluating differential cross sections derived from the signal sample and comparing
them to MINERvA's published results for the semi-inclusive channels 
$\numu$-CC($\pi^{+}$) and $\anumu$-CC($\pi^{0}$)~\cite{Carrie-pion}.
In contrast to the $\pi^+ p$ hadronic systems of $\numu$-CC($\pi^{+}$),
the total amplitude for reaction \eqref{exclusive-channel} 
receives contributions from the I = 1/2 isospin amplitude, $A_{1}^{CC}$,
as well as from the I = 3/2 isospin amplitude, $A_{3}^{CC}$:
\begin{equation}
\label{eq:Isospin_Amp}
 A(\nu n\rightarrow\mu^{-}p\pi^{0})=A^{CC}(p\pi^{0})=\frac{2}{3}(A_{3}^{CC}-A_{1}^{CC}).
\end{equation}
The amplitudes associated with neutrino-induced baryon-resonances 
having the same spin and the same orbital angular momentum may interfere.
The consequence for $\pi^{0} p$ final states is that many more 
interference terms are possible for their transitions $\left|A^{CC}(p\pi^{0})\right|^{2}$
than occur in $\left|A^{CC}(p\pi^{+})\right|^{2}$~\cite{Rein-Sehgal}.

\subsection{CC($\pi^{0}$) measurements and phenomenology}

Differential cross sections for exclusive-channel CC($\pi$) interactions 
including reaction \eqref{exclusive-channel}, were obtained 
during the 1970s and 1980s era of large bubble chamber experiments 
and the measurements continue to be of interest at the present time.
Neutrino-induced single-pion production at $E_{\nu} \leq 1.5$\,GeV 
was studied using hydrogen and deuterium fills in  
Argonne National Laboratory\textquoteright s (ANL) 
12-ft diameter bubble chamber;  cross sections and distributions of
pion-nucleon invariant mass and of $Q^{2}$ were reported~\cite{Barish-1976, Radecky-1982}.
Similar measurements with higher statistics were subsequently obtained using
Brookhaven National Laboratory\textquoteright s (BNL) 7-ft deuterium-filled
bubble chamber exposed to a neutrino beam with an average energy of 1.6 GeV~\cite{Kitagaki-1986}.
In these two experiments, the $\Delta^{++}(1232)$ was found to dominate the $p\pi^{+}$ final-state with
no other resonance structure observed.   
For the $p\pi^{0}$ and $n\pi^{+}$ final states however, $\Delta^{+}$ production
was accompanied by a broad $N\pi$-mass distribution extending to 2.5 GeV.
Measurements of CC($\pi$) reactions at much higher $E_{\nu}$, 
for $\anumu$ as well as for $\numu$ scattering,
were reported in the late 1980's by experiments using the SKAT heavy liquid bubble chamber 
at Serpukhov ($ 3  <  E_{\nu} < 30\,$GeV)~\cite{SKAT-1989}, and by the deuterium-filled 
Big European Bubble Chamber at CERN
($\langle E_{\nu} \rangle = 54$ GeV)~\cite{Allasia-1990}.    
The neutrino-induced $\pi^{0} p$ invariant mass distributions 
of the latter experiments showed the contributions from higher-mass baryon resonances 
to be increasing relative to the $\Delta(1232)$ 
contribution in this final state, and in the $\pi^{+} n$ final state as well.

For CC($\pi^{0}$) production on hydrocarbon targets, there are only 
two previous measurements.   These were carried out relatively recently
using neutrino beams with flux maxima occurring well below 2.0 GeV.  
The MiniBooNE experiment used its spherical, 12\,m diameter 
\v{C}erenkov detector filled with mineral oil (CH\textsubscript{2}) 
exposed to neutrinos of 0.5 to 2.0 GeV with peak flux at 0.6 GeV. 
The CC($\pi^{0}$) cross section was obtained, and 
differential cross sections for muon and $\pi^{0}$ kinematics and for 
$Q^2$ were reported~\cite{MiniBooNE-pi0-2011}.   The
K2K experiment used neutrino scattering on extruded scintillator (polystyrene) bars 
to obtain the ratio of CC($\pi^{0}$) to quasielastic scattering at a mean
neutrino energy of 1.3 GeV~\cite{K2K-2011}.

Recent theoretical treatments of 
CC($\pi^{0}$) can be roughly categorized in terms of eras of endeavor.   
In the pre-MiniBooNE era, investigations focussed on refining the phenomenology and
used the bubble chamber data to obtain cross checks 
on the formalism~\cite{Mosel-PRC73-2006, Paschos-2nd-res-region, 
Nieves-PRD76-2007, Mosel-PRC-2009, Mosel-PRD-2010}.
With the advent of CC($\pi^{+,0}$) measurements 
reported by MiniBooNE~\cite{MiniBooNE-piplus-2011, MiniBooNE-pi0-2011},
the focus shifted to developing explanations for 
the new data~\cite{Mosel-PRC87-2013, Nieves-PRD87-2013, Review-2014}.
Comparisons between predictions and MiniBooNE results turned up 
discrepancies in the shapes of distributions of final-state $\pi^{0}$ mesons,
highlighting the interplay between pion FSI processes 
and possible formation time effects~\cite{Review-2014}.   Comparisons with the
MiniBooNE data have continued into the present era, 
however phenomenological studies have expanded their purview
to the higher-energy pion production results reported by MINERvA 
and to measurements underway by T2K~\cite{Mosel-PRC91-2015, Mosel-PRC88-2013,
Mosel-PRD89-2014, Yu-Paschos-PRD-2015}.

\subsection{CC($\pi^{0}$) measurement using MINERvA}
The fine-grained tracking capability of the MINERvA experiment's 
central scintillator tracker coupled with the MINOS downstream spectrometer are
used to investigate a sample of 6110 events having the topology and kinematics of
the signal channel \eqref{signal-channel}.   In contrast to previous measurements, this
investigation covers the neutrino energy range that is directly relevant to the 
NOvA and DUNE neutrino oscillation programs.   Results are obtained with good statistics on a 
hydrocarbon target, a medium whose atomic number composition 
is an excellent match to NOvA's liquid scintillator medium
and does not differ greatly from T2K's water medium.   
This analysis obtains differential cross sections for channel~\eqref{signal-channel} 
that characterize the kinematics of both the final-state
muon and the produced $\pi^{0}$.    
The results reported here complement and extend
MINERvA's previous measurements of $\numu$ and $\anumu$ CC pion production 
on hydrocarbon~\cite{Brandon-pion, Trung-pion, Carrie-pion}.   

The analysis makes use of MINERvAÕs fine-grained tracking to undertake
measurements that heretofore have never been carried out for neutrino scattering
on a hydrocarbon medium.   For a subsample of channel-\eqref{signal-channel} events, 
a leading final-state proton has been reconstructed in conjunction with the $\pi^{0}$, 
enabling the final-state hadronic invariant mass to be directly measured.   

A further selection on hadronic invariant mass yields a subsample 
confined to the region of $\Delta(1232)^+$ production.  It is
used to examine the decay angular distribution 
of the $\pi^{0} p$ system in the candidate $\Delta(1232)^+$ rest frame.
In bubble chamber experiments of
the 1970s and 1980s, polarization effects were observed in the pure I = 3/2 channel 
$\numu p \rightarrow \mu^- p \pi^+$~\cite{Radecky-1982, Kitagaki-1986, Allen-NP-1986, Allasia-1990} and in the 
mixed isospin channel $\anumu p \rightarrow \mu^+ p \pi^-$~\cite{Allen-NP-1986}.
This work reports the first-ever measurement of the decay polar and azimuthal angles 
$\theta$ and $\phi$ for the mixed isospin reaction~\eqref{exclusive-channel}.   
The latter angular distributions are potentially complicated 
since nearby resonances can interfere strongly 
with the leading $\Delta^+$ amplitude~\cite{Rein-Sehgal, Rein-Z-1987}.


\section{Overview of Data and Analysis}
\subsection{Detector, Exposure, and $\nu$ Flux}
\label{subsec:B-D-E}
MINERvA uses a fine-grained, plastic scintillator tracking 
detector~\cite{minerva-NIM-2014, minerva-NIM-2015} positioned upstream of the magnetized 
MINOS near detector~\cite{minos-NIM-2008}, to record interactions 
of neutrinos from the high-intensity NuMI beam 
at Fermilab~\cite{NuMI-Beam-2016}. 
In the present analysis the spectrometer's central, scintillator tracking 
region serves as the target, with the surrounding electromagnetic and hadronic 
calorimeters providing containment.   The fiducial volume has a hexagonal 
cross section of 2.0\,m minimal diameter, extends longitudinally for 
2.4\,m, and has a mass of 5400\,kg.  It consists of 106 planes, each of $\sim2$\,cm thickness,
composed of polystyrene scintillator strips oriented transversely to the 
detector's horizontal axis.   The planes are configured into modules.  There are two planes per
module, with an air gap of $2.5$\,mm between each module.
The horizontal axis is inclined 
3.3$^\circ$  relative to the beam axis.   Three scintillator-plane 
orientations, at 0$^\circ$ and $\pm 60^\circ$  relative to the detector 
vertical axis, provide  X, U, and V ``views'' of interactions in 
the scintillator medium.  The planes alternate between UX and 
VX pairs,  enabling 3-D reconstruction of vertices, charged tracks, and 
electromagnetic showers of neutrino events.   The downstream electromagnetic calorimeter (ECAL)
is identical to the central tracking region but includes a 0.2\,cm (0.35 radiation lengths) lead sheet in front
of every two planes of scintillator.  The readout electronics have a timing resolution 
of 3.0\,ns for hits of minimum-ionizing particles~\cite{Marshall-2016},  allowing 
the efficient separation of multiple interactions within a  
single 10\,$\mu$s beam spill.

The MINOS near detector is located 2\,m downstream of MINERvA and serves as the muon 
spectrometer for MINERvA's central tracker.  A muon that exits the downstream surface
of MINERvA is tracked by the magnetized, steel-plus-scintillator planes 
of MINOS, and the muon's momentum and charge are measured.  
Trajectories of individual muons traversing the two detectors are matched together
by correlating the positions, angles, and timings of track segments in each detector.

The data were taken between March 2010 and April 2012 with the NuMI beamline operating
in a mode that produces a wide-band neutrino flux whose spectrum peaks at 3.0 GeV,  
extends from 1 GeV to above 20 GeV, and has $\numu$ content at 93\% purity.
The event sample analyzed here is obtained with an integrated 
exposure of $3.04 \times 10^{20}$ protons on target (POT).

The $\numu$ flux is calculated using a detailed simulation
of the NuMI beamline based on GEANT4~\cite{Geant4-2003, Allison-2006} (version 9.2.p03, physics list FTFP\_BERT).
The flux simulation is constrained using proton-carbon yield 
measurements~\cite{Alt-NA49-2007, Barton-PRD-1983, Lebedev-Thesis} together with
more recent thin-target data on hadron 
yields~\cite{NuMI-Flux-Aliaga-2016}.   A further constraint on the flux estimation
is derived using the $\nu + e^{-}$ scattering rate observed by \minerva \cite{Park-PRD-2016}.

\subsection{Neutrino interaction modeling}
\label{subsec:II-B}
The reference Monte Carlo (MC) simulation used by this analysis is based upon
version 2.8.4 of the GENIE neutrino event generator~\cite{Andreopoulos-NIM-2010}.    
The GENIE stratagies that underwrite the generation of CC neutrino-nucleus
interactions in the simulation are the same as used in previous MINERvA studies of CC($\pi$) 
and are described in MINERvA publications~\cite{Brandon-pion, Carrie-pion}.
Recent neutrino measurements and phenomenology developments motivate certain augmentations to the GENIE version;
these are implemented using event re-weighting and by adding a simulated sample of quasielastic-like $2p2h$ events.
All of the augmentations (described below) have been used in recent \minerva works~\cite{Ren-PRD-2017, Betancourt-2017}.

In brief, the target nucleus is modeled as a relativistic Fermi gas with addition of a high-momentum tail 
required to account for short-range correlations~\cite{Bodek-Ritchie-1981}.
Neutrino-induced pion production arises from interaction with a single nucleon and 
proceeds either by baryon-resonance excitation or by non-resonant DIS processes.    
Baryon-resonance pion production is simulated using the Rein-Sehgal model~\cite{Rein-Sehgal-1981} 
with incorporation of modern baryon resonance properties~\cite{PDG-2012}.
Decays of baryon resonances are generated isotropically in their rest frames, but an exception is made for
the $\Delta^{++}$.   In MINERvA's GENIE-based simulation, $\Delta^{++}$ decays are generated with
an angular isotropy at 50\% of the strength predicted by Rein-Sehgal~\cite{Brandon-pion}.
Non-resonant pion production is modeled according to the formalism of Bodek-Yang~\cite{Bodek-2005-PS} 
with parameters adjusted to reproduce electron and neutrino scattering measurements 
over the invariant hadronic mass range $W < 1.7$~GeV~\cite{Gallagher-2006,Wilkinson-PRD-2014, Rodrigues-EurPhys-2016}.
No allowance is made for interferences among baryon-resonance amplitudes.

The simulation includes a treatment of final-state intranuclear interactions (FSI) of pions and nucleons that
are created and propagate within the struck nucleus.   Accounting for pion FSI is especially important to this
analysis because of the large pion-nucleon cross sections that occur in the vicinity of 
$\Delta$-resonance excitation.  In GENIE, an effective model 
for FSI is used in lieu of a full intranuclear cascade;  each pion is allowed to have at most
one rescattering interaction while traversing the nucleus~\cite{Dytman-2011-CP}.  This approximation works
well for a light nucleus such as carbon and it makes the simulation amenable to event reweighting.
It has been shown by \minerva studies of $\numu$-CC($\pi^{+}$) and $\anumu$-CC($\pi^{0}$) that GENIE FSI
improves the agreement between simulation and data 
for pion kinematic distributions~\cite{Trung-pion, Brandon-pion, Carrie-pion}.

Events are added to the GENIE-based simulation to include quasielastic-like $2p2h$ interactions whose generation is based 
on the Valencia model~\cite{Nieves-PLB-2012, Gran-PRD-2013}, but with the interaction rate
raised to match the data rate observed in MINERvA inclusive CC scattering data~\cite{Rodriques-2p2h-2016}.
For quasielastic scattering, kinematic distortions attributed to long-range nucleon-nucleon correlations are included in 
accordance with the Random Phase Approximation (RPA) calculation of Ref.~\cite{Nieves-PRC-2004}.
For CC single pion production, the GENIE prediction 
for the nonresonant pion contribution has been reduced by 53\% for all pion charge states,
as this has been shown to yield a better agreement 
with the deuterium bubble chamber data~\cite{Wilkinson-PRD-2014, Rodrigues-EurPhys-2016}.

\subsection{Detector response, calibrations, event isolation}
The ionization response of the \minerva spectrometer to muons, electrons and hadrons  
is modeled using GEANT4~\cite{Geant4-2003, Allison-2006} (9.4.p02, QGSP\_BERT). 
The scale for visible energy is established by requiring agreement 
between data and simulation for the reconstructed
energy deposited by through-going muons that are momentum-analyzed using MINOS.
The scale for muon dE/dx energy loss in the detector 
is known to within 2\%.   Reconstruction of the energy of hadronic showers makes use of calorimetric corrections.
Initially these were estimated from simulation according to procedures detailed in Ref.~\cite{minerva-NIM-2014}.
Subsequently the corrections were refined and validated using measurements obtained with a scaled-downed replicate of
the detector operated in a low-energy particle test beam~\cite{minerva-NIM-2015}.   The test beam data, in
conjunction with in-situ measurements, underwrite the determinations of tracking efficiencies and energy
responses to protons, charged pions, and electrons, and of the value assigned to the Birks' constant of the
scintillator.

The central scintillator tracker has a radiation length 
of 40\,cm which corresponds to 25 planes
for photons traveling normal to the planes.   
The energies of photon and electron-induced showers are 
reconstructed by calorimetry using calibration constants 
determined from the simulation~\cite{Trung-pion, Park-PRD-2016, Wolcott-PRL-2016}.
For electromagnetic (EM) showers with visible energies above 700 MeV, 
the conversion factor from visible energy, $E^{\gamma}_{visible}$, 
to true energy, $E^{\gamma}_{true}$, is constant to good approximation.
The EM showers of this analysis however come from 
$\pi^{0} \rightarrow \gamma \gamma$ decay and have energies in the range of 50 MeV to 1 GeV.
Over the lower half of this range, Compton scattering 
competes with pair production in the total photon-carbon cross section, and so a conversion
factor $(E^{\gamma}_{true}/E^{\gamma}_{visible})$ is used that increases  
linearly, from 1.33 to 1.49, as $E^{\gamma}_{visible}$ is decreased from 700\,MeV to 0\,MeV~\cite{Altinok-Thesis-2017}.

\smallskip

For each 10\,$\mu$s NuMI beam spill, the visible energy in the scintillator 
as a function of time is divided into ``time slices" of tens to hundreds of nanoseconds; 
each time slice encompasses a single event in the detector.   A charged particle
initiated by a neutrino interaction traverses the scintillator strips of the central tracker, and
its trajectory is recorded as individual energy deposits (hits) 
having a specific charge content and time-of-occurrence.   The hits are grouped in time, and
neighboring hits in each scintillator plane are gathered into objects called clusters.
Clusters that have more than 1\,MeV of energy are matched among the three views to create a
track.   The position resolution per plane is 2.7\,mm and the track
angular resolution is better than 10\,mrad~\cite{minerva-NIM-2014} in each view.

\section{Event reconstruction and selections} 
\subsection{Muons, protons, and vertex energy}
\label{subsec:mu-p-vtx-energy}
A track that starts in the central ``tracker",  
exits via the detector's downstream surface,
and matches with a negatively-charged track that enters the
front of MINOS near detector, is taken to be the $\mu^-$ track of a CC event.
Selected events are required to have a muon track that is MINOS-matched.
This requirement eliminates events having muons 
with production angle $\theta_{\mu}>25$ degrees ($\approx 26\%$ of CC interactions in the tracker).
The momentum resolution for muons 
reconstructed using both the \minerva and MINOS detectors (the RMS width
of the residual fractional error) is $6.1\%$.  A small inefficiency occurs with muon track reconstruction
due to event pileup which requires a correction of $-4.4\%$ ($-1.1\%$) to the simulated efficiency for muons of 
momenta less than (greater than) 3 GeV/c~\cite{Carrie-pion}.

Having found and reconstructed a muon track, the reconstruction algorithm searches for additional tracks that share
the primary vertex of the muon track.   If additional shorter track(s) are found, the vertex position is refitted. 
Kinked tracks, which are usually the result of secondary interactions, are then reconstructed by
searching for additional tracks starting at the endpoints of tracks previously found.
The event primary vertex is required to occur within the central 112 planes of the scintillator tracking region and must be
at least 25\,cm from any edge of the planes.  The fiducial volume contains a target mass of 5.27 metric tons
with 3.17 $\times 10^{30}$ nucleons.  

Events having no reconstructed tracks emanating from the 
primary vertex other than the muon are retained for further analysis.
When one or more extra tracks are found, selections 
for protons are applied.   The ionization dE/dx profile of each track
is compared to simulated profiles for protons and charged pions 
based on the Bethe-Bloch formula, and a likelihood ratio is calculated.
An event is retained if all prompt non-muon tracks pass a cut 
on the proton consistency score~\cite{Walton-PRD-2015}; the momenta
of such tracks is then assigned according to track range for the proton hypothesis.
To further ensure that final-states having $\pi^+$ mesons are eliminated, 
regions around the primary vertex, around each track endpoint,
and around beginning and endpoints of candidate EM showers 
that are remote from the primary vertex, are examined for 
instances of Michel electrons.   These are low-energy 
($ \le100$ MeV) EM showers that arise in the decay sequence 
$\pi^{+}\rightarrow\mu^{+}\rightarrow e^{+}$ and appear 
in time slices that are later, by 0.5 to 16\,$\mu$s, than the candidate-event time.
Events associated with a Michel tag are removed.

Protons of kinetic energy $T_{p} \simeq 100$ MeV are at the threshold for reconstruction as tracks in the detector.   
A candidate for channel \eqref{signal-channel} can have one or more final-state 
protons that are too short to be tracked, but still gives visible ionization around the primary vertex.   This ``vertex energy"
appears as hits that are not used in reconstruction of the muon or prompt proton tracks.    The unused hits that fall within a 
sphere of 9\,cm radius centered on the primary vertex are gathered and their net energy estimated using a calibration established
via study of inclusive hadron production.    

\begin{figure}
\begin{centering}
\includegraphics[scale=0.21]{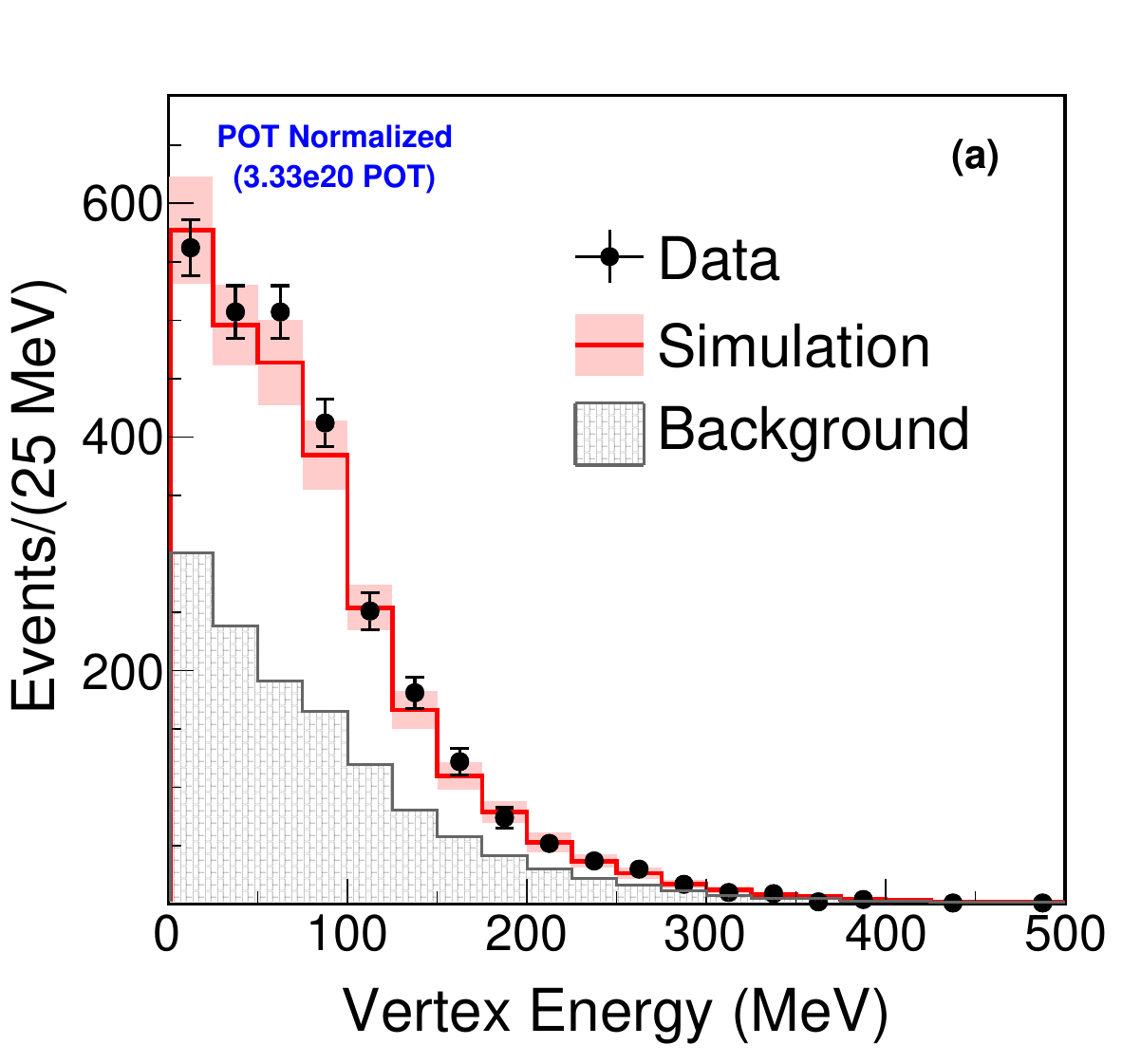}
\includegraphics[scale=0.21]{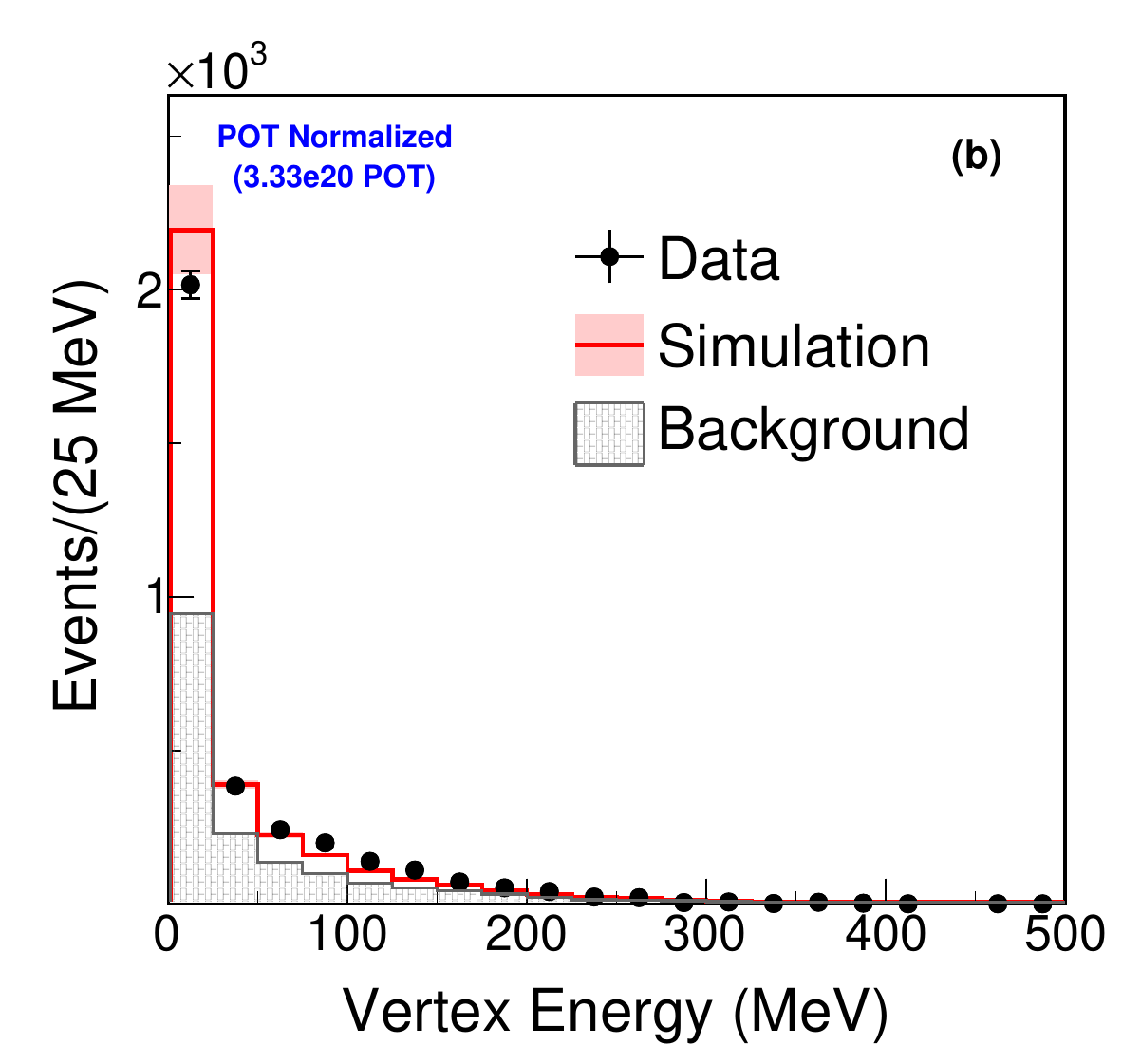}
\par\end{centering}
\caption{Vertex energy in data and simulation, of events after all selections.  Statistical (systematic) errors are shown for 
the data (for the MC prediction).  The simulation
is broken out as signal (clear histograms) and background (shaded histograms).
The distribution shapes depend
on whether the muon is accompanied by zero reconstructed protons (a), or by
one or more reconstructed protons (b).}
\label{Fig01ab}
\end{figure}

The distributions of vertex energy for channel \eqref{signal-channel} candidates after
all analysis selections, including the $\pi^{0}$ selections described in the remainder of this Section, 
are shown in Fig.~\ref{Fig01ab} together with predictions from the GENIE-based MC.  The background
consists of events that pass the signal selection criteria but have final states that, upon emergence from their parent nuclei,
are not examples of reaction \eqref{signal-channel}.

Vertex energy consists of ionizations of low-momentum protons (or charged pions) from the primary vertex that cannot be 
reconstructed as tracks.    When proton(s) of an event are reconstructed as tracks, their energy is excluded from
``vertex energy".   Thus the energy reach of the vertex-energy distribution for events without reconstructed protons (Fig.~\ref{Fig01ab}a)
is noticeably reduced for the subsample that possesses reconstructed protons (Fig.~\ref{Fig01ab}b).   However for each of 
the two subsamples which together comprise the analysis signal sample, the vertex energy distribution 
is well-described by the simulation.

\subsection{Selections for events with two photons}

The candidate sample is subjected to a pre-filter prior to $\pi^{0}$ reconstruction.
An event is removed if the upstream nuclear targets section has registered time-coincident
energy that exceeds 20 MeV.    Such activity can be due to interactions
originating in a nuclear target or in the earth berm upstream of the detector.
Events are removed if the unused visible energy (hits not associated with muons, 
primary protons or vertex energy) in the detector (central tracker plus ECAL
and HCAL) is less than 50 MeV -- effectively too little to yield a reconstructable $\pi^{0}$.
Events are also removed if the unused visible
energy in the detector exceeds 2500 MeV since a single, produced $\pi^{0}$ is highly unlikely
to have that much energy.   The signal loss incurred by these two cuts is less than 1\%.

Events selected to this stage are subjected to $\pi^{0}$ pattern recognition and reconstruction.
The goal of pattern recognition is to identify events that have two and only two gamma
showers that belong to the decays of singly produced neutral pions, $\pi^{0} \rightarrow \gamma \gamma$.
For each event, the unused hit clusters are gathered that are coincident to within 25 ns with the muon track,
have ionization that exceeds low-activity cross-talk, and lie within the scintillator or the ECAL regions.
Hit clusters found in the X view that are close in polar angle with respect to
the vertex (the best-fit origin of the muon track plus any proton tracks), but can be separated in radial distance from the vertex, 
are grouped into photon-conversion candidates. 
Then, for each candidate, clusters in the U and V views 
that are consistent among the three views are added. 
Photon candidates must have clusters in at least two views 
in order to enable their directions to be reconstructed in three dimensions.   
Additional steps are taken when the above procedures yield a single-photon or three-photon
configuration.   Searches are repeated using tighter polar angle criteria, and using the U view and V view
instead of the X view for the start-of-search.   For three-photon situations, a spurious shower can sometimes
be identified on the basis of a straight-line fit to the positions of each cluster plus the event vertex.
These extra steps are estimated to have resolved the candidacy of $9.6\%$ of
the final signal sample~\cite{Altinok-Thesis-2017}. 

The photon reconstruction is then finalized for events that are deemed to have exactly two photon showers.
The position, direction, and energy of a photon-conversion shower are determined by
the clusters that have been assigned to each of the candidate photons. 
The photon direction is reconstructed from the cluster energy-weighted slopes in each view.
The photon vertex is defined using the closest cluster to the event
vertex on the photon direction axis.  The photon energy is reconstructed 
by calorimetry using calibration constants determined by detector response simulations. 
An event is removed if the conversion distance for the more energetic photon (denoted as $\gamma_{1}$)
is less than 14\,cm ($\sim$ 0.28 conversion lengths) from the primary vertex.   This cut is based on optimization of sample purity; it
mitigates against non-tracked charged particle hits close to the vertex that can fake an EM shower.

\begin{figure}
\begin{centering}
\includegraphics[scale=0.30]{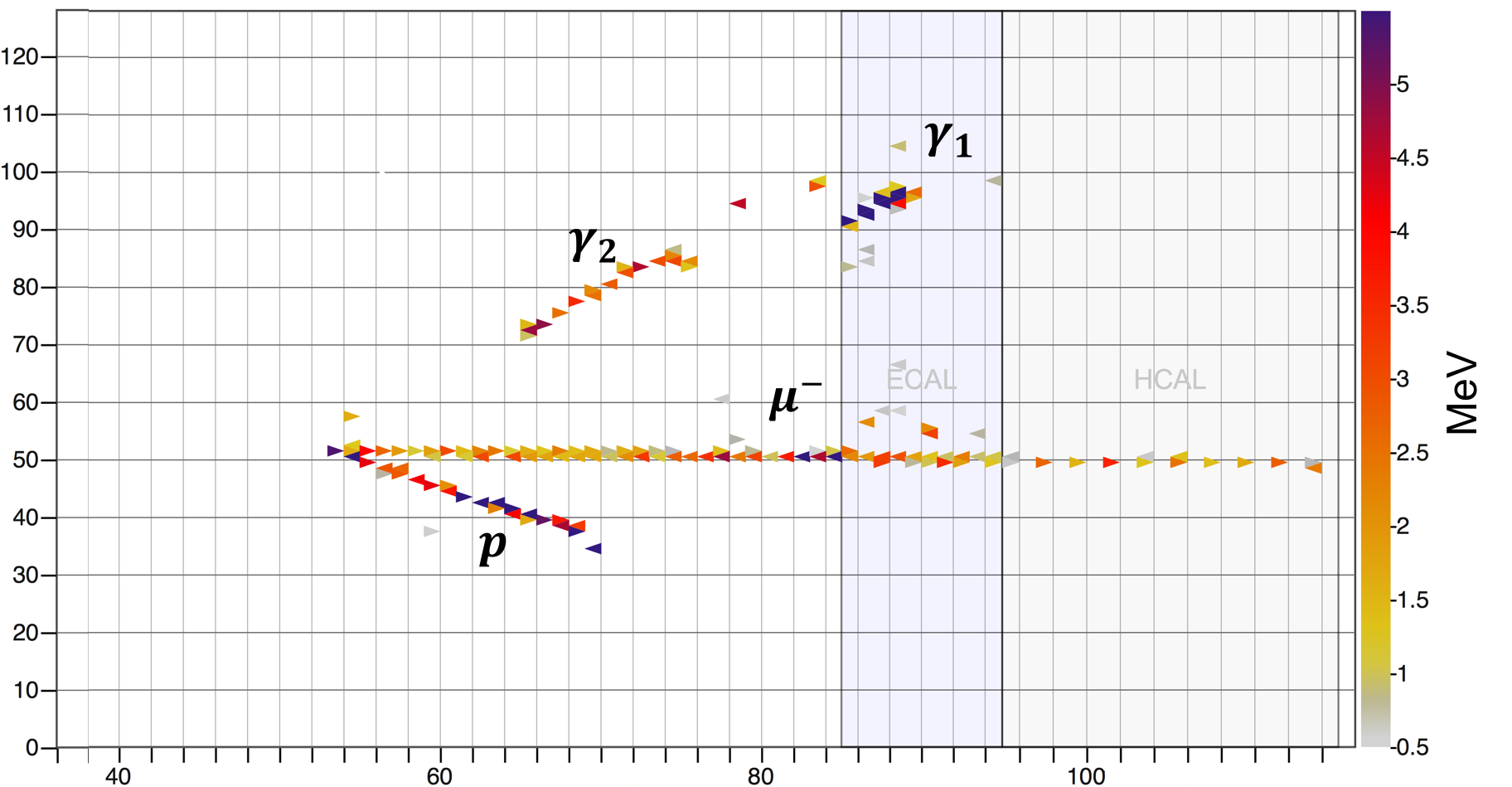}
\caption{Data event candidate for subsample \eqref{exclusive-channel} 
of the signal channel \eqref{signal-channel}. 
The neutrino enters from the left and interacts in the scintillator 
to produce a muon, proton, and two photon conversion showers. 
The horizontal and vertical axes labels show the module 
and strip numbers, respectively.  The color (online) linear scale shows energy deposited in the strips.} 
\label{Fig02}
\par\end{centering}
\end{figure}

Figure~\ref{Fig02} shows a data event whose topology satisfies the requirements
for retention in the signal sample.     Clearly discernible are the final-state
muon, the pair of photon conversion showers (one in scintillator, the other in the ECAL), and a 
proton that ranges to stopping.  The interaction vertex is nearly devoid of extra ionization hits. 
The final-state muon traverses the scintillator, ECAL, and HCAL
regions and exits in the direction of the MINOS near detector.  The photon converted in the ECAL, $\gamma_{1}$, 
is more energetic (618 MeV) than the photon $\gamma_{2}$ in the scintillator (140 MeV).  The 
display is obtained with the Arachne event viewer~\cite{Arachne}.   The photons and proton of this event are  
energetic compared to selected-sample norms, but the uncluttered appearance of the event topology is
typical of the sample.

\smallskip

\subsection{The $\pi^{0}$ signal}
\label{sec:pi0-reco}
The photon energies $E_{\gamma 1}$, $E_{\gamma 2}$ and their opening angle from the vertex, $\theta_{\gamma\gamma}$,
are used to calculate the two-photon invariant mass $M_{\gamma\gamma}$:
\begin{equation}
M_{\gamma\gamma}^2 = 2E_{\gamma_{1}}E_{\gamma_{1}} (1-\cos\theta_{\gamma\gamma}). 
\label{eq:invmass}
\end{equation}

\begin{figure}
\begin{centering}
\includegraphics[scale=0.39]{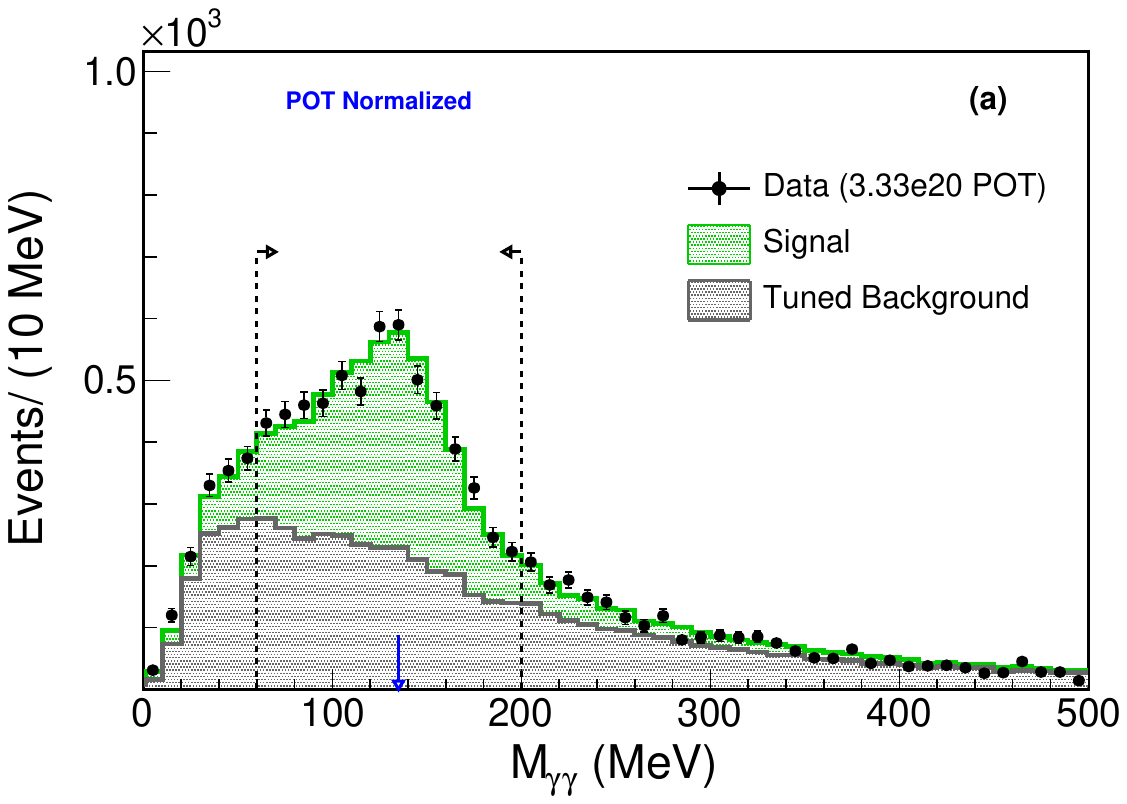}
\includegraphics[scale=0.39]{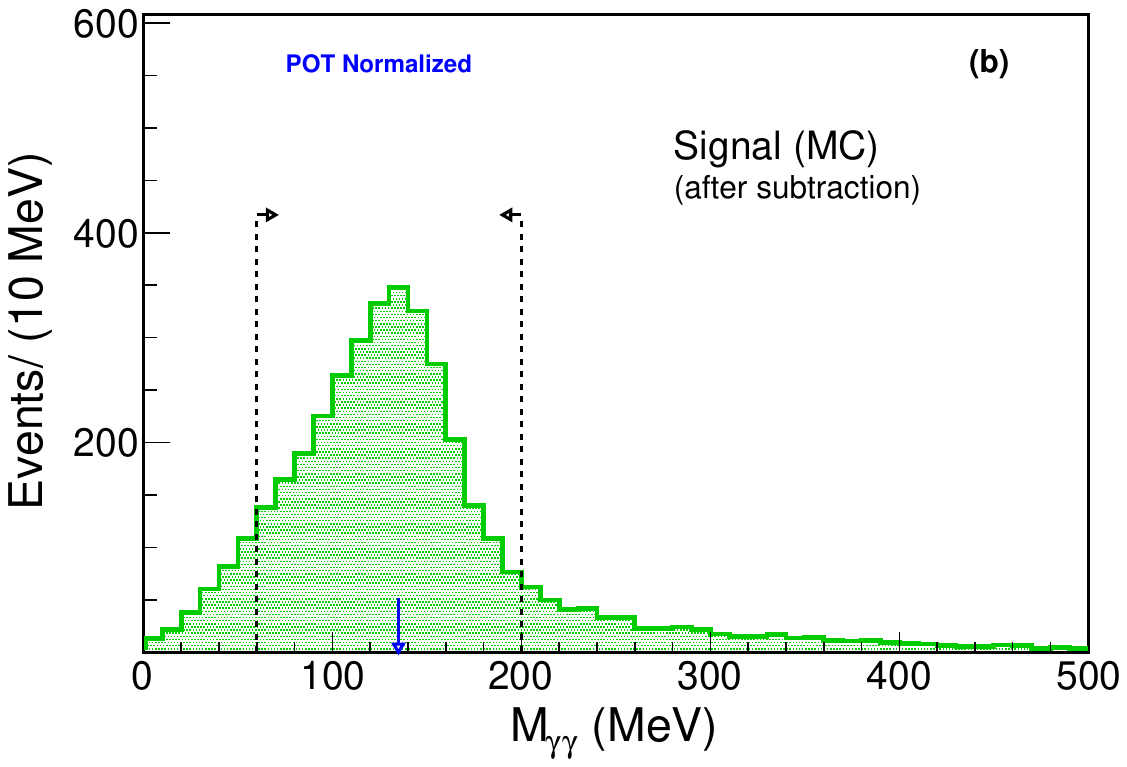}
\par\end{centering}
\caption{The $M_{\gamma\gamma}$ distribution from data (solid circles in (a)) 
and as predicted by the MC (histograms, both plots).  In (a), the MC prediction
for signal-plus-background is seen to be in good agreement with the data.
In (b), the signal distribution as estimated by the MC is replotted. It
distributes broadly about the $\pi^{0}$ mass and exhibits a skew towards
lower mass values (b).}
\label{Fig03}
\end{figure}
\smallskip

\noindent
The invariant mass $M_{\gamma\gamma}$ is studied in the data and in the MC simulation.   The data distribution is examined 
after background subtraction, where the background consists of selected events whose out-of-nucleus final-state particle content
is not consistent with reaction \eqref{signal-channel}.
The background estimation is constrained by fitting the data to four different sideband distributions
(see Sec.~\ref{sec:Side-Band-Fit}).   The peak position for the data distribution was found to be $137.8\pm2.6\textrm{ MeV}$,
while the peak position for the MC distribution was $130.3\pm1.6\textrm{ MeV}$.   
Correction factors were then applied to the 
absolute scales for electromagnetic energy, separately in the data and the MC, 
that adjusted the $M_{\gamma\gamma}$ peaks to match the 
nominal $\pi^{0}$ mass $(134.97\textrm{ MeV})$.   These tunings of EM absolute energy scales have been applied to
the distributions of Fig.~\ref{Fig03} and to all subsequent Figures.

Figure~\ref{Fig03} shows the $M_{\gamma\gamma}$ distribution for the data (solid circles), 
for the signal distribution as predicted by the MC (upper histogram in Fig.~\ref{Fig03}a and histogram of Fig.~\ref{Fig03}b),
and for the background distribution predicted by the MC (lower histogram in Fig.~\ref{Fig03}a).
The distributions shown are normalized to the POT of the data exposure, with the normalization of the estimated
background (Fig.~\ref{Fig03}a, lower histogram) constrained 
via fitting to four data sidebands as described in Sec.~\ref{sec:Side-Band-Fit}.   The peak in the data 
coincides with the $\pi^{0}$ mass.   The signal-plus-background as estimated by the MC follows the shape of the data (Fig.~\ref{Fig03}a, upper
histogram).   The estimated signal, which is shown in Fig.~\ref{Fig03}b after background subtraction,
distributes broadly and skews toward lower invariant mass.   The skew reflects the loss of hits from EM showers that travel transversely
to the detector's longitudinal axis.

\subsection{Estimation of $E_{\nu}$, $Q^{2}$, and $W$} 
\label{subsec:Estimate-E-Q-W}
The neutrino energy is calculated as the sum over the energies 
of all reconstructed final-state particles, plus the
vertex energy (Sec.~\ref{subsec:mu-p-vtx-energy}), 
plus additional extra energy within 
the event time slice that is associated with 
the event and is not included in the particle reconstructions: 
\begin{equation}
E_{\nu}=E_{\mu}+E_{\pi^{0}}+\sum T_{p}+E_{vtx}+E_{extra}.
\label{eq:neutrino-energy-estimate}
\end{equation}
Here, $E_{\pi^{0}}$ is assembled from the gamma vector momenta 
and the $\pi^{0}$ rest mass (see Sec.~\ref{sec:Pion-Kin}),
$T_{p}$ is the kinetic energy of a reconstructed proton, $E_{vtx}$
is the vertex energy, and $E_{extra}$ is the extra energy 
that is left over after reconstruction and is not included in $E_{vtx}$.  

\begin{figure}
\begin{centering}
\includegraphics[scale=0.21]{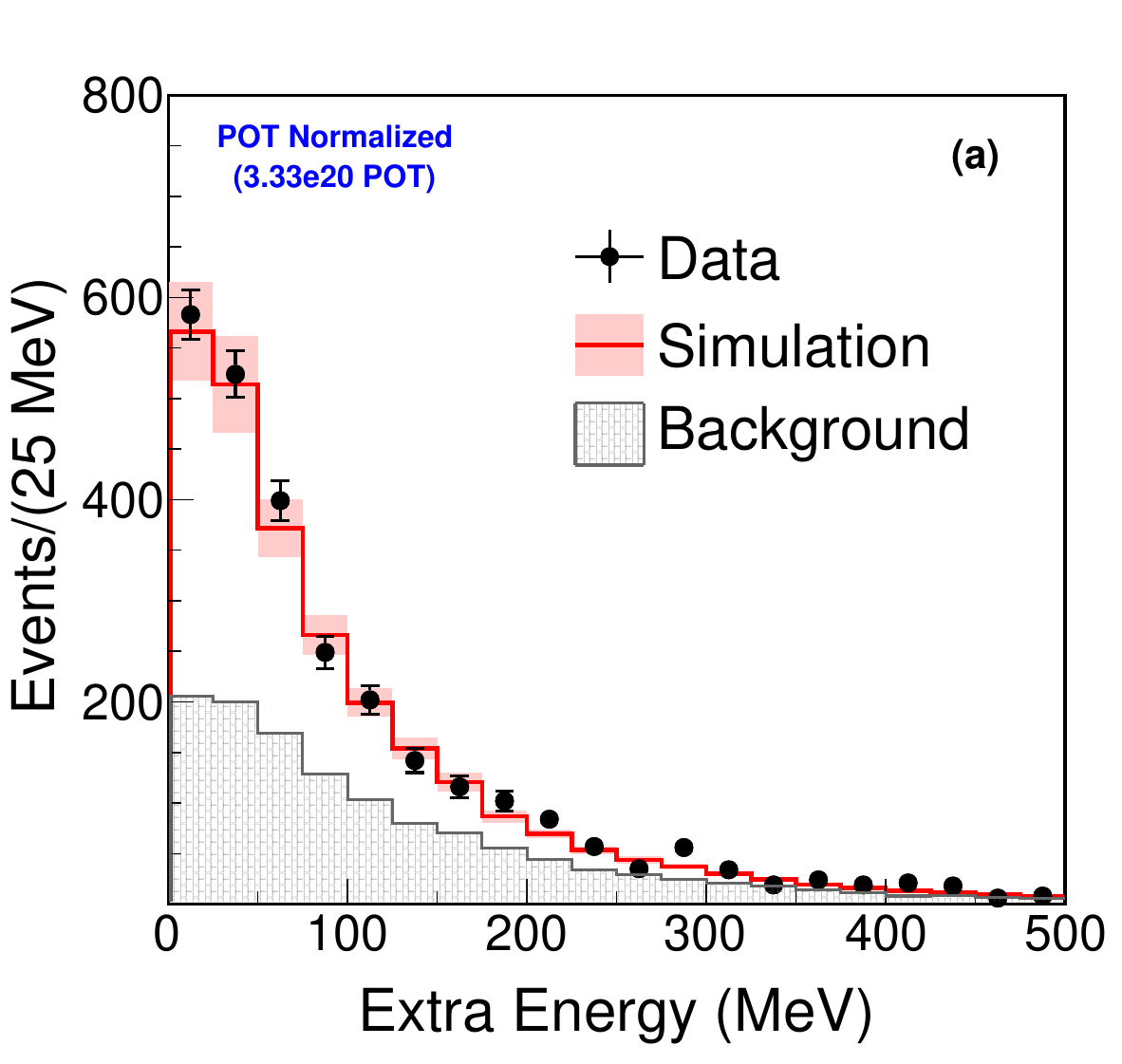}
\includegraphics[scale=0.21]{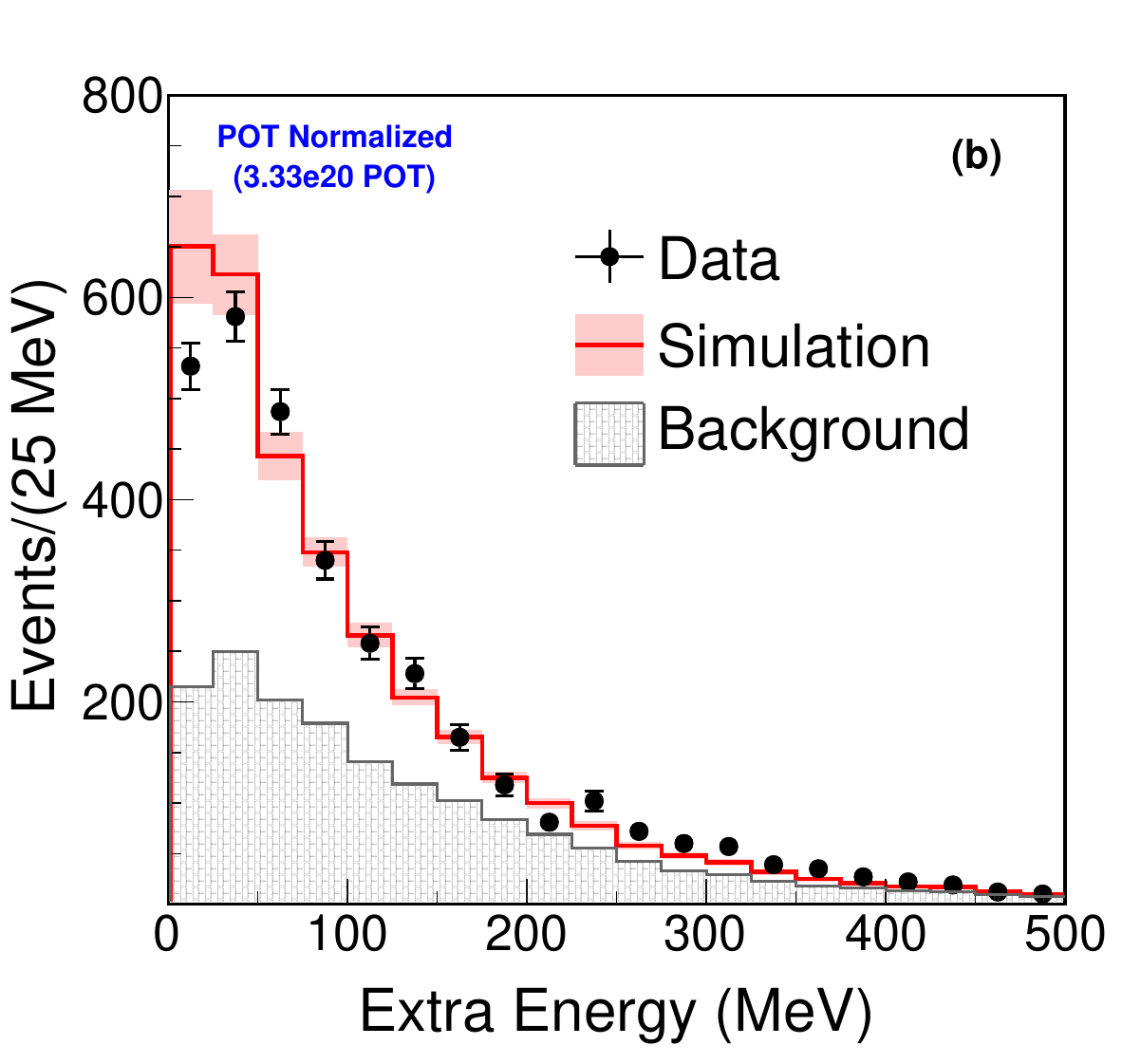}
\par\end{centering}
\caption{Distributions of event energy $E_{extra}$ that is
included in the neutrino energy estimation -- see text.   The distributions are 
for events without a reconstructed proton (a), and for 
events with at least one reconstructed proton (b). }
\label{Fig04}
\end{figure}

Extra energy is the sum of energies represented by all unused hit clusters that are time-coincident with the muon track.  The hits include
clusters rejected during $\pi^{0}$ reconstruction, clusters lying close to the muon but not used by the tracking algorithm, and all 
unused clusters that lie within a radius of 30\,cm about the primary vertex.  Also included in $E_{extra}$ are ionizations remote
from the primary vertex that are induced by scatters of slow neutrons released by breakup of the struck nucleus.
Figure~\ref{Fig04} shows the energy $E_{extra}$
that is included in the neutrino estimation, for events without 
(Fig.~\ref{Fig04}a)  and with (Fig.~\ref{Fig04}b) reconstructed protons.
The distributions have similar shapes and are adequately described by the GENIE-based simulation.    

\begin{figure}
\begin{centering}
\includegraphics[scale=0.21]{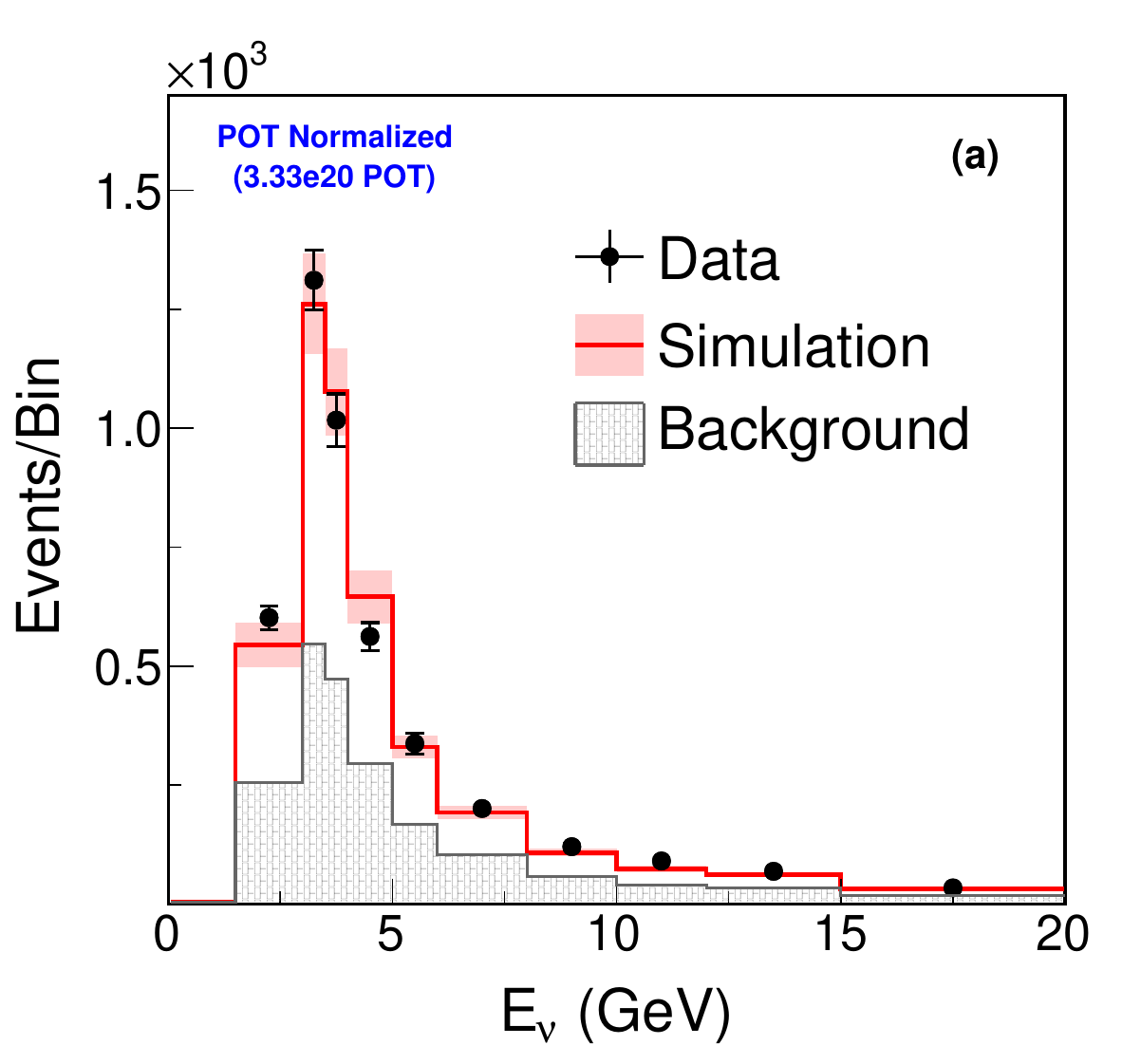}
\includegraphics[scale=0.21]{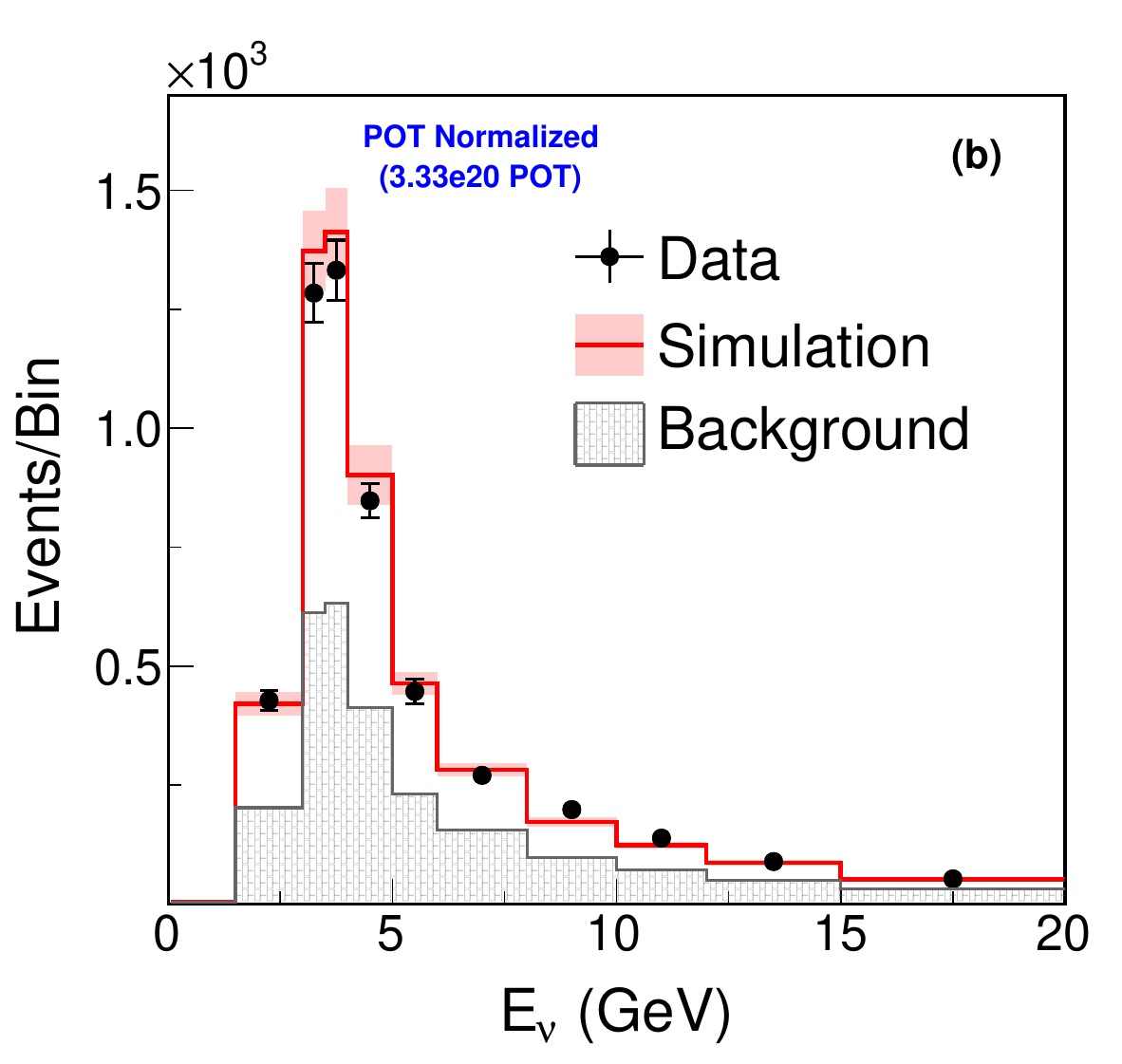}
\par\end{centering}
\caption{Neutrino energy distributions for selected events that do not have, or do have,  
reconstructed proton(s) in the final state (plots (a) and (b) respectively).}
\label{Fig05}
\end{figure}

The sum of final-state energies as in Eq.~\eqref{eq:neutrino-energy-estimate} is used to reconstruct the neutrino energy, $E_{\nu}$.
Distributions of reconstructed-$E_{\nu}$ minus true-$E_{\nu}$ as evaluated by the reference MC are found to be symmetric and
have means consistent with zero~\cite{Altinok-Thesis-2017}.    This $E_{\nu}$ estimator is used throughout the analysis for both the data and simulation.
Figure~ \ref{Fig05} shows the distributions of calculated $E_{\nu}$ for
events without a reconstructed proton (Fig.~\ref{Fig05}a) and for events that have at least one reconstructed proton (Fig.~\ref{Fig05}b).   
The errors shown with the data points are purely statistical.   The simulation predictions are in rough agreement with the data for
both distributions.  The neutrino energy distribution for
the full signal sample has an average neutrino energy of 5.25 GeV 
(4.4 GeV for events with $E_{\nu}$ below 10 GeV).  The RMS width of the MC (reconstructed Ð true)
distribution for $E_{\nu}$ is 0.42 GeV.

The four-momentum-transfer-squared, 
$Q^{2}$, and the hadronic invariant mass, $W$, are calculated as follows:
\begin{equation}
\label{def-Q2}
Q^2 = -(k-k')^2 =  2E_\nu(E_\mu-|\vec{p}_\mu|\cos\theta_{\mu})-m_\mu^2 ,
\end{equation}
and
\begin{equation}
\label{def-W}
 W^2 = (p+q)^2 = M_N^2 + 2M_N(E_\nu-E_\mu)- Q^2 ,
\end{equation}
where $k$, $k'$, and $p$ are the four-momenta of the incident neutrino, the
outgoing muon, and the struck nucleon respectively, while
$q=k-k'$ is the four-momentum transfer and $M_N$ is the nucleon mass. 
For estimation of $W$ as in Eq.~\eqref{def-W} there is an underlying assumption
that a single struck nucleon is initially at rest.   It is useful at this point to distinguish between
$W_{exp}$ and the ``true W" of the simulation.  The hadronic mass, $W_{exp}$, is calculated -- in both data and
simulation -- using observable quantities.   The variable $W_{exp}$ can be estimated 
for every signal event using reconstructed lepton kinematic variables 
and calculating with Eqs.~\eqref{def-Q2} and \eqref{def-W}.  
This can be done whether or not a final-state proton was reconstructed.
The RMS widths of MC (reconstructed - true) distributions for the variables $Q^2$ and $W_{exp}$ 
are 0.02 GeV$^2$ and 0.09 GeV respectively and both distributions are sharply peaked at zero~\cite{Altinok-Thesis-2017}.

\subsection{Final selections: the signal sample} 
\label{subsec:Final-select}
Selections that finalize the signal sample of the analysis are now imposed.
As previously stated, the muon polar angle with respect to the beam axis is required to be $< 25^{\circ}$.
Additionally the reconstructed neutrino energy is limited to $1.5\textrm{ GeV}<E_{\nu}<20\textrm{ GeV}$.
The lower bound ensures good acceptance for muons to be matched in MINOS, and the
upper bound removes events that are unlikely to be CC single-$\pi^{0}$ production.   
An upper bound of 1.8 GeV is placed on $W_{exp}$ in order to obtain a sample that is 
enriched in baryon resonance events.
Finally, it is required that $M_{\gamma\gamma}$, calculated according to \eqref{eq:invmass}, lies in the range 
$60\textrm{ MeV}<M_{\gamma\gamma}<200\textrm{ MeV}$.     The signal region around the nominal $\pi^{0}$ mass
is shown by the pair of vertical delimiters in Fig.~\ref{Fig03}.

The final selected signal sample contains 6110 data events. The purity of the signal 
sample is calculated using events that have a vertex inside
the fiducial volume and a muon track reconstructed using the MINOS
near detector. The sample purity is 51\% implying that 3115 of the 
selected data events are actual occurrences of Eq.~\eqref{signal-channel}. 
The reconstruction efficiency for signal events is 8.4\%.   The requirement
that the muon of an otherwise valid signal event gives a MINOS-matched track,
accounts for an efficiency loss of nearly 51\%.

Background reactions remaining in the selected signal sample
are classified into three categories having sizable statistics, plus an ``other" category
consisting of an odd-lot of strange particle production, CC anti-neutrino events, and neutral current events.   
The largest background category consists of events that have zero $\pi^{0}$s but one or more
charged mesons emerging from the target nucleus (with $\pi^+$ being most probable); 
this ``charged meson(s)" category gives 57\% of the background.  Of the produced charged mesons in this category
approximately 37\% subsequently interact in the detector and initiate $\pi^{0}$s. 
Other mesons of this category scatter and/or travel transversely in the detector, giving hit clusters that mimic gamma conversions.
Backgrounds also arise from events that have at least one $\pi^{0}$ 
that emerges from the struck nucleus, accompanied by additional pions or kaons.   
This ``$\pi^{0}$+meson(s)" category accounts for 20\% of 
background events.  A third background arises with events that have zero mesons, but have proton and neutron-induced 
ionizations that give a fake $\pi^{0}$ in reconstruction.   This ``zero meson" category accounts for 19\% of 
backgrounds; the remaining 4\% of background is the ``other" contribution.

\section{Background rates from fitting to sidebands}
\label{sec:Side-Band-Fit}
The overall normalizations for the background categories are
determined by fitting the MC simulation to multiple different background-rich samples 
whose events have topological and kinematic resemblances to the selected signal events.
The data events that populate these ``sideband samples" do not appear 
in the selected signal sample as they do not satisfy one or more of the event selections.
The individual sidebands have different mixtures of the background categories 
so that a combined fit has sensitivity to the normalizations of all categories.

\begin{figure}
\begin{centering}
\includegraphics[scale=0.38]{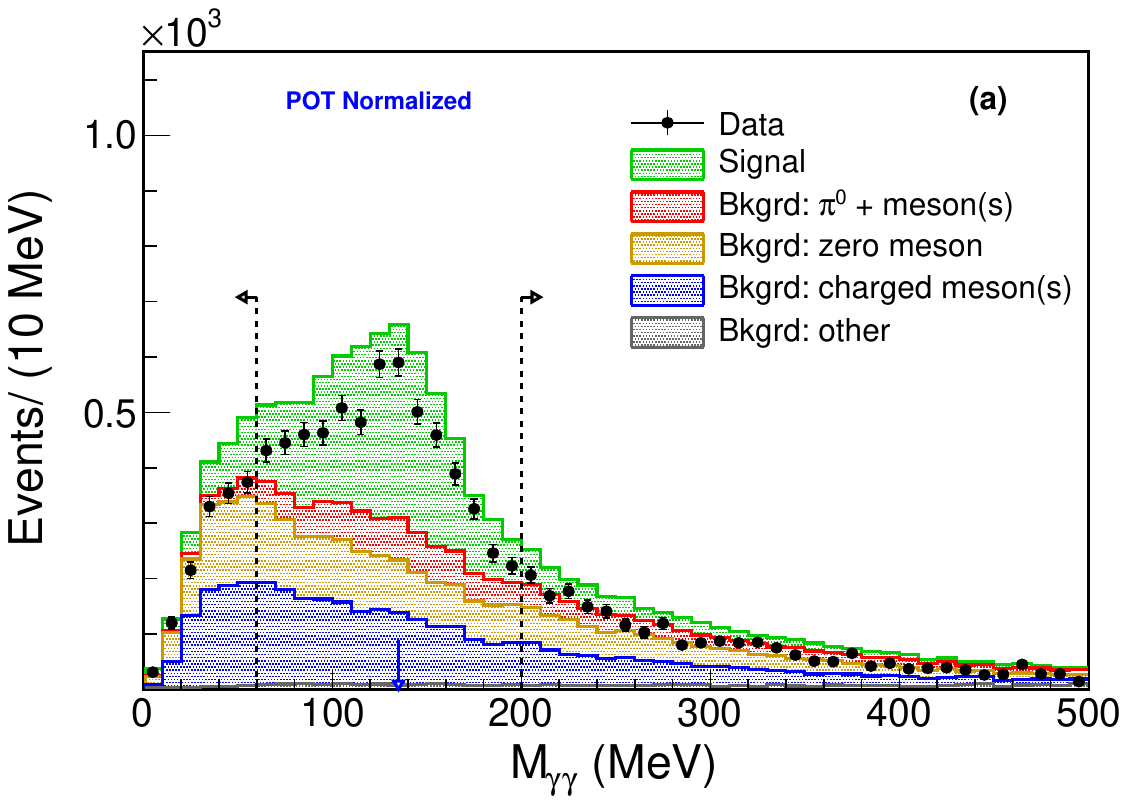}
\includegraphics[scale=0.38]{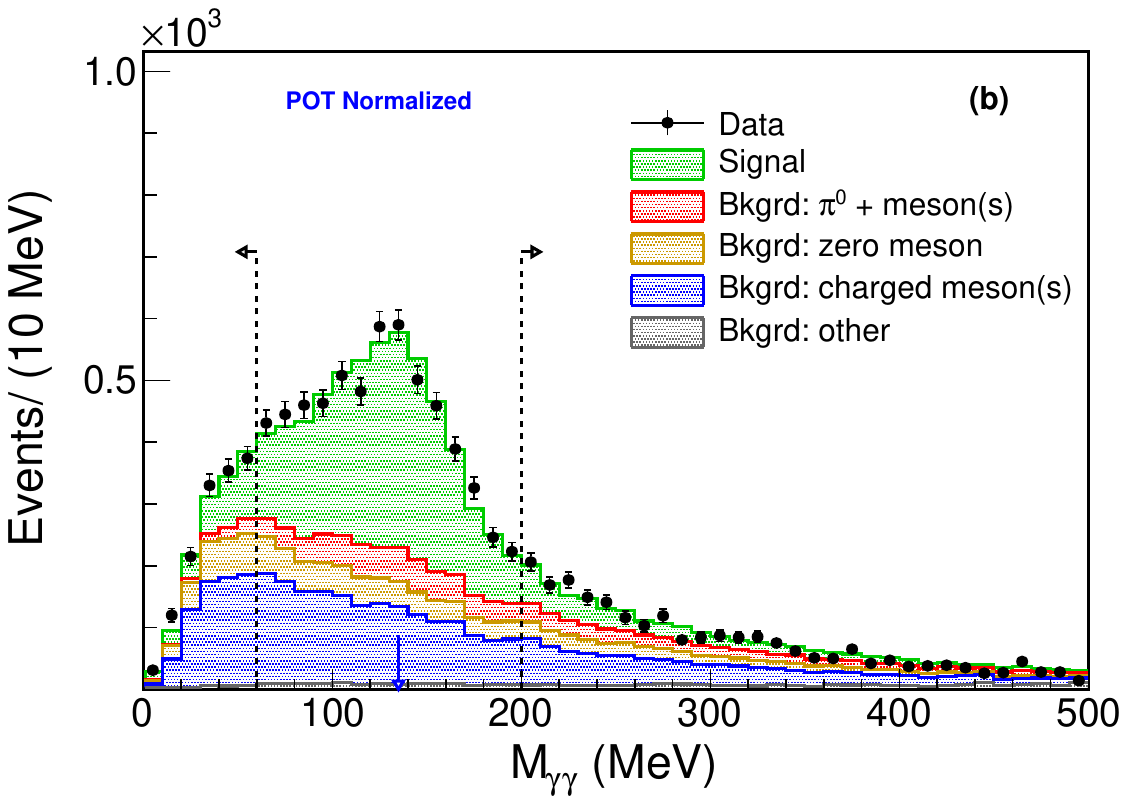}
\par\end{centering}
\caption{Data versus MC $\gamma\gamma$ invariant mass
distributions before (a) and after (b) revision of the background normalizations to match the values obtained
by the overall fit to the four data sideband samples.   The low-side and high-side $M_{\gamma \gamma}$ sidebands,
denoted by arrows in the plot, are fitted together with the Michel-tag and low-proton-score sidebands, while the data of 
the signal region between 60 MeV and 200 MeV are excluded.   The MC agreement with the data in the signal region
improves dramatically as the result of constraining the backgrounds in the sideband regions.}
\label{Fig06}
\end{figure}

The analysis uses four separate sideband samples to achieve good constraints on the
normalizations of the three main background categories~\cite{Altinok-Thesis-2017}.
The first two sidebands consist of events whose $M_{\gamma \gamma}$ values fall below
or above the $\pi^{0}$ invariant mass selection indicated by the 
vertical bars displayed in Figs.~\ref{Fig03} and 
\ref{Fig06}.   Figure~\ref{Fig06} shows that
for the low-side $M_{\gamma \gamma}$ sideband, 
the MC simulation -- before fitting -- has the zero-meson and
charged-meson(s) background categories contributing at comparable rates, 
with the $\pi^{0}$+meson(s) category giving a much smaller contribution.
For the high-side $M_{\gamma \gamma}$ sideband, zero-meson and charged-meson(s) are still the
leading categories, however the $\pi^{0}$+meson(s) category has a larger presence.   
The $M_{\gamma \gamma}$ low-side and high-side sidebands contain
1424 and 2309 data events respectively.

\begin{figure}
\begin{centering}
\includegraphics[scale=0.38]{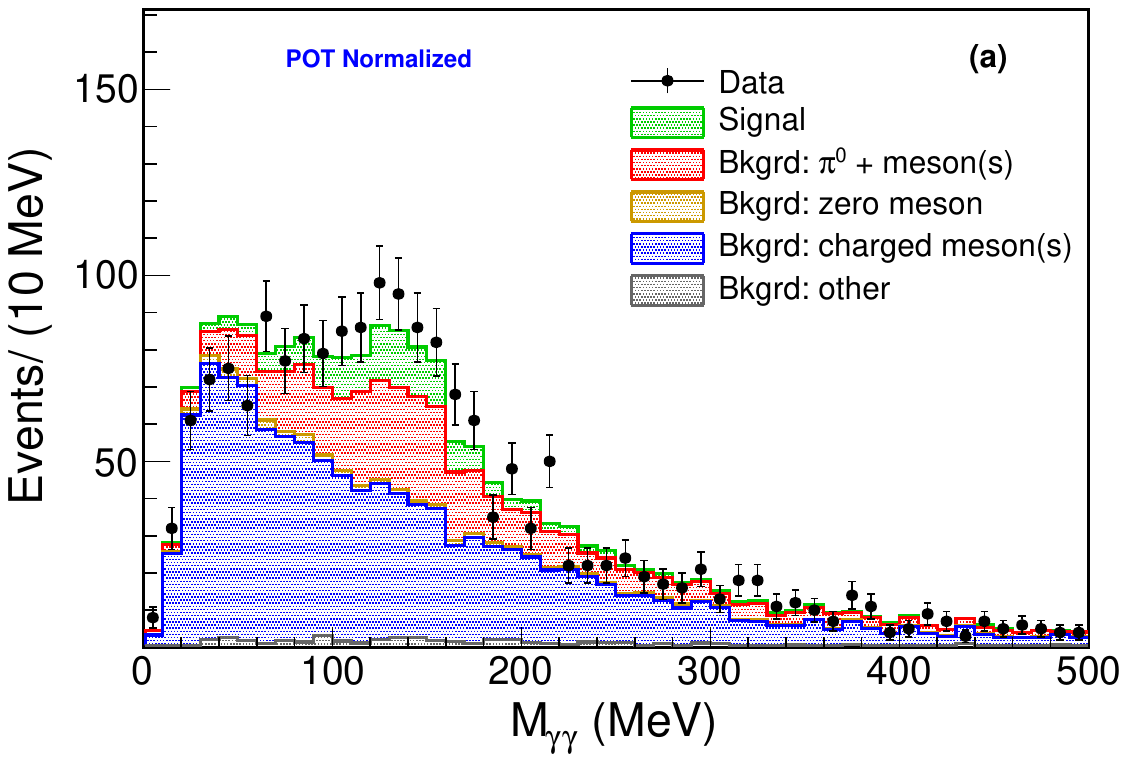}
\includegraphics[scale=0.38]{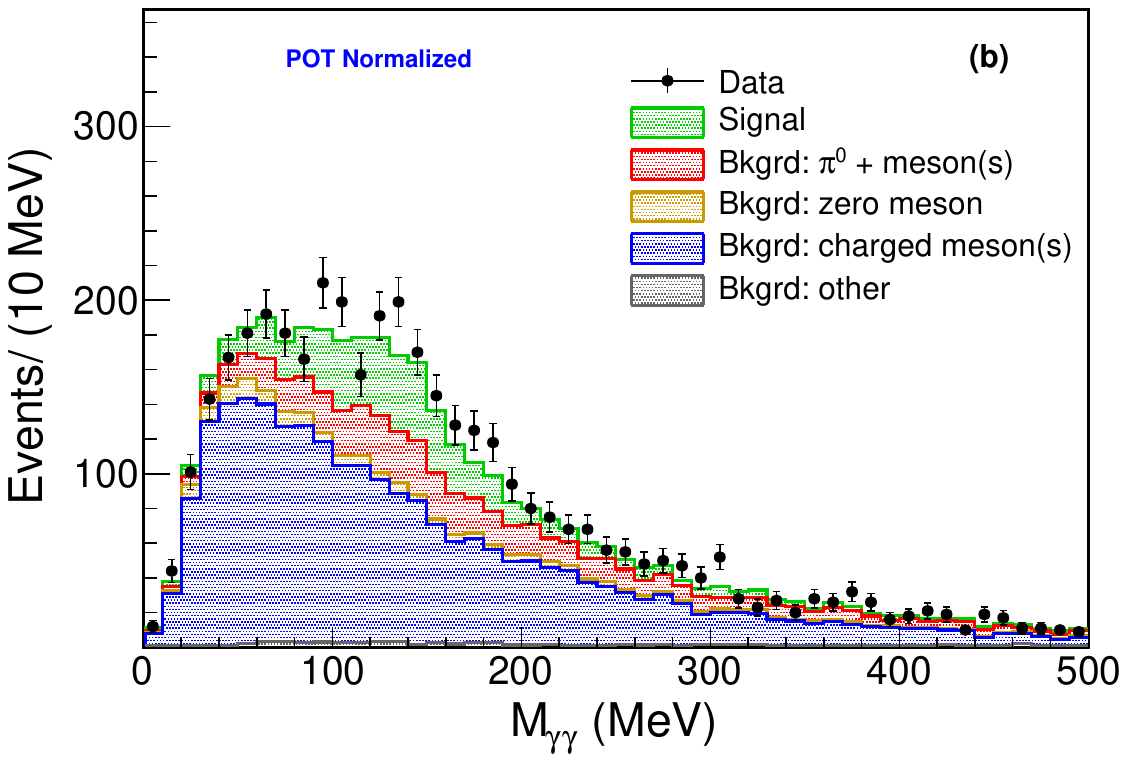}
\par\end{centering}
\caption{Comparisons of $M_{\gamma \gamma}$ distributions of data (solid points) to predictions of the reference MC (histograms),
for the Michel-tag sideband (Fig.~\ref{Fig07}a) and for the low-proton-score sideband (Fig.~\ref{Fig07}b).   The MC predictions are brought into
agreement with the data by adjustments of component background levels as determined by the simultaneous fit to all four sidebands of the analysis.}
\label{Fig07}
\end{figure}

A third sideband contains events tagged as having 
Michel EM showers from endpoint $\pi^+$ decays.
This Michel sideband (1803 events) has abundant charged-meson(s) 
background but also contains a sizable $\pi^{0}$+meson(s) contribution.
There are very few zero-meson events in the Michel sideband.
In the fourth sideband, the muon is accompanied 
by a second reconstructed track which has a low
likelihood score for the proton hypothesis (3933 events).   
This low-proton-score sideband is also predicted to be mostly composed of charged meson(s)
plus $\pi^{0}$+meson(s) backgrounds but with their apportionment differing somewhat from that
in the Michel-tag sideband.    The estimated compositions of the latter sidebands after background
tuning by the fit-to-sidebands described below, are displayed as component histograms in Figs.~\ref{Fig07}a and
\ref{Fig07}b.

The fit to the sidebands is carried out as a single chi-squared fit 
of the simulation prediction to the data, where the data consists
of binned values of $\gamma \gamma$ invariant masses taken over all all four sidebands.  
The fit procedure varies the normalizations of the charged-meson(s), 
zero-meson, and $\pi^{0}$+meson(s) categories 
to minimize the total $\chi^{2}$ in all sidebands at once.  
The MC estimates of other background and of residual signal content
of the sidebands are held fixed.   The $\chi^{2}$ uses 50 bins 
for each of the Michel-tag and low-proton-score sidebands, 30 bins
for the high-side $M_{\gamma \gamma}$ sideband, 
and 6 bins for the low-side $M_{\gamma \gamma}$ sideband.   
The total of 136 bins, each with good data statistics, 
yields robust constraints on the background normalizations.   
The fit improves the $\chi^{2}$  in each of the four sidebands.   The $\chi^{2}$ over all sidebands changes 
from 956.3 to 216.8 for 133 degrees of freedom.
  
Figure~\ref{Fig06} shows the $\gamma\gamma$ invariant mass distribution for the 
data compared to the GENIE MC prediction before and after the sideband fit.
The sideband fit uses the low-side and high-side $M_{\gamma \gamma}$
regions (designated via arrows), but not the signal region between 60 MeV and 200 MeV.   

The $\chi^{2}$ for the entire distribution, with 49 degrees of freedom, changes from 656.6 (Fig.~\ref{Fig06}a) to 88.2 (Fig.~\ref{Fig06}b)).  
Figure \ref{Fig07}a,b shows that the sideband fit also brings the MC background model into agreement with the data distributions
of the Michel-tag and low-proton-score sidebands.
The normalization scale factors obtained by the fit are 0.97$\pm$0.03 for the charged-meson(s) background
category, 0.72$\pm$0.07 for the $\pi^{0}$+meson(s) category, and 0.42$\pm$0.04 for the zero-meson category.  
The fit tends to reduce the zero-meson category in all four sideband samples~\cite{Altinok-Thesis-2017}.
This trend may indicate that the generation and/or subsequent visible scatters of neutrons is overpredicted 
in the simulation for these quasielastic-like zero-meson events~\cite{Elkins-Masters-2017}.

\section{Determination of cross sections}
\label{X-sec-calc}
Calculation of the flux-integrated differential cross section per nucleon
for kinematic variable $X$ (such as $p_{\mu}$, $\theta_\mu$, and $Q^{2}$), 
in bins of $i$, proceeds as follows~\cite{Brandon-pion,Trung-pion, Carrie-pion}:
\begin{equation}
\label{eq:dif-xsec}
( \frac{d\sigma}{dX} )_{i} =  \frac{1}{T_{n}\Phi } \frac{1}{\Delta X_i} 
	\frac{\sum\limits_{j} U_{ij} ( N^{data}_{j} - N^{bkg}_{j} ) }{\epsilon_{i}},
\end{equation}
where $T_{n}$ is the number of nucleons in the fiducial volume,
$\Phi$ is the integrated flux, $\Delta X_i$ is the bin width,
$\epsilon_{i}$ is the selection efficiency and acceptance.   The  
unfolding function, $U_{ij}$, calculates the contribution
to true bin $i$ from reconstructed bin $j$, where the $j$th bin contains
$N_{j}^{data}$ data candidates and $N_{j}^{bkg}$ of estimated background events.
Calculation of $\sigma(E_{\nu})_{i}$, the cross section 
per bin $i$ of neutrino energy, is carried out using
an expression that can be obtained from 
Eq.~\eqref{eq:dif-xsec} by dropping $\Delta X_i$ and changing $\Phi$ to
$\Phi_{i}$, the neutrino flux for the $i$th bin of $E_{\nu}$.

The factor $(N^{data}_{j} - N^{bkg}_{j})$ in Eq.~\eqref{eq:dif-xsec} denotes the binned contents of the 
background-subtracted data distribution, obtained by subtracting 
the MC background prediction after sideband tuning 
from the data.  The background-subtracted data is subjected 
to an iterative unfolding procedure~\cite{D'Agostini-NIM-1995}.  
The procedure takes detector resolution smearing into account 
and corrects reconstructed values to true values according to mappings, $U_{ij}$, determined by the reference simulation.  
For most of the kinematic variables measured in this work, the unfolding matrices are close to diagonal and
the effects of unfolding are minor.    The largest spread about the matrix-diagonal is exhibited by
pion kinetic energy; the matrices for $Q^2$ and $E_{\nu}$ also show some population in neighboring off-diagonal bins.
For all variables, the biases versus the number of unfolding iterations were studied.   The biases, after the four iterations that were used,
were always found to be much smaller than other systematic uncertainties.

The bin-by-bin efficiency $\epsilon_{i}$ is estimated using the simulation.
For muon momentum, for example, the efficiency rises from 3\% below 2 GeV/c to 10\% at 3.5 GeV/c and then remains roughly
constant, reflecting the limited tracking acceptance ($\theta_{\mu} < 25^{\circ}$) for lower-momentum muons 
in the MINOS near detector.   As previously stated, the overall reconstruction
efficiency for signal events is 8.4\%.

The analysis uses current determinations of the integrated and differential neutrino fluxes over the $E_{\nu}$ range 1.5 to 20\,GeV 
for the NuMI low-energy beam mode~\cite{NuMI-Flux-Aliaga-2016}.  The neutrino flux in bins of $E_{\nu}$ is given in the Supplement~\cite{Supplement}.
The value for the integrated flux $\Phi$ is 2.55$\times10^{-8}$ $\numu$/cm$^2$/POT.   Also required is the number of target nucleons inside the 
fiducial volume: $T_{n} = 3.17 \times 10^{30}$ nucleons.

\section{Systematic Uncertainties}
The cross section measurements use the reference simulation to estimate selection efficiencies, detector acceptance and resolutions,
distribution shapes of backgrounds, and the neutrino flux.   All of these estimations introduce systematic uncertainties. 
While there are many individual sources of uncertainty, each can be associated with 
one of five general categories.   Detector response uncertainties arising 
with particle energy scales, particle tracking and detector composition 
comprise the first category.   Categories two, three, and four are,
respectively, uncertainties with simulation modeling of neutrino interactions, 
uncertainties residing with the GENIE model for FSI involving produced hadrons, 
and neutrino flux uncertainties.   A fifth category is reserved for ``other" uncertainties 
which include uncertainties with background fitting and subtraction, uncertainty 
inherent to the unfolding procedure, and uncertainty arising from inclusion of $2p2h$ events into the simulation.   

Most sources of uncertainty for the present work were encountered 
by previous  \minerva studies of CC($\pi)$ interactions and their treatment 
is described in publications~\cite{Brandon-pion, Trung-pion, Carrie-pion}.   
The systematic uncertainty from the neutrino flux is described in detail 
in Refs.~\cite{NuMI-Flux-Aliaga-2016, Fields-PRL-2013}.    
The systematic uncertainties for many quantities are evaluated by shifting the relevant parameter in the simulation
within its $\pm 1\sigma$ band and producing a new simulated event sample.   Cross sections are then remeasured using an ensemble of such 
alternate-reality samples, and a covariance matrix is formed from the results.   The procedure is repeated for each systematic source; 
details are given in~\cite{Brandon-pion}.   On cross-section plots to follow, the errors shown represent the square roots of covariance diagonal entries.  
The full correlation matrices are given in the Supplement~\cite{Supplement}.

\begin{figure}
\begin{centering}
\includegraphics[scale=0.43]{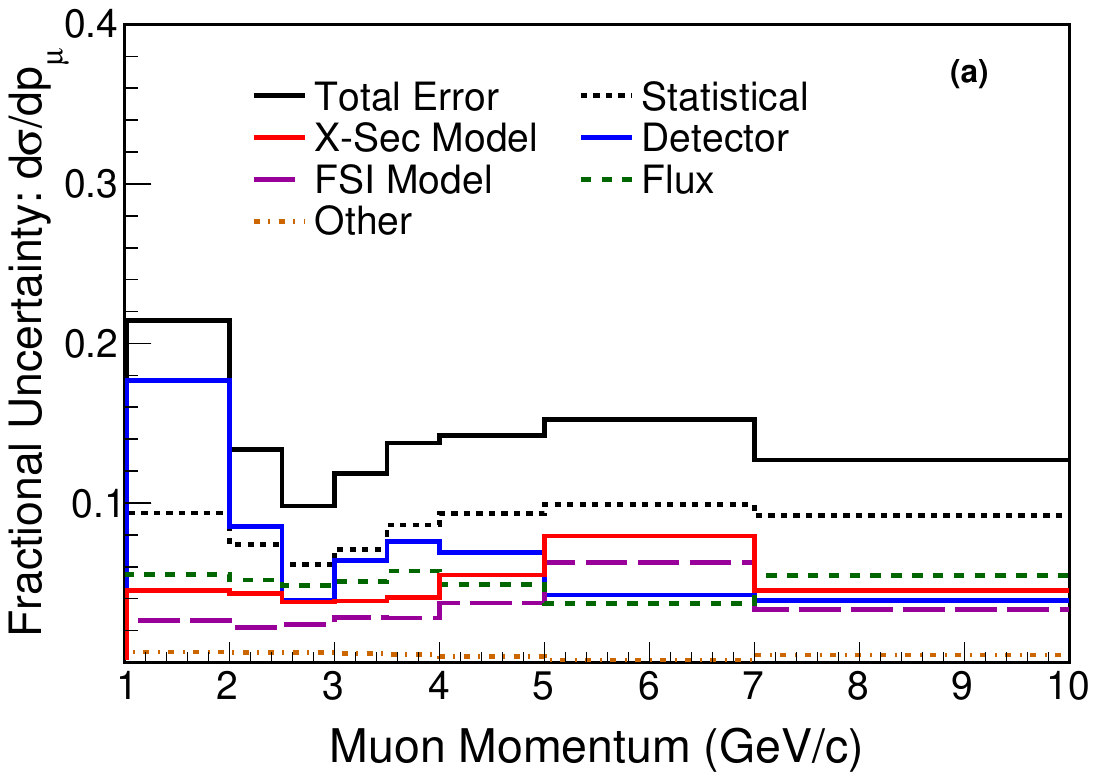}
\includegraphics[scale=0.43]{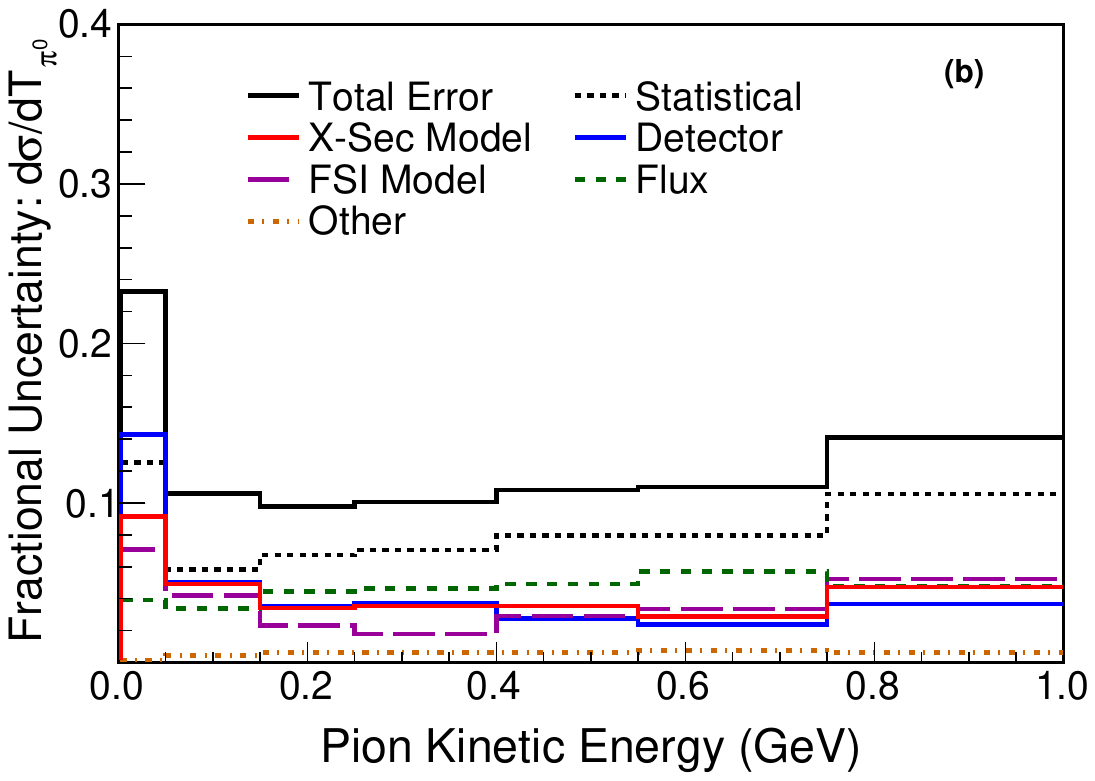}
\par\end{centering}
\caption{Systematic error composition of the total fractional uncertainties for
the differential cross sections of muon momentum (a) and of $\pi^{0}$ kinetic energy (b).
Component histograms show the statistical errors (dotted line)and the contributions from the 
five systematic-error categories (see text).   Detector response (blue solid-line histograms) gives the leading uncertainty in the lowest
bins of either distribution, while statistical uncertainty is leading in higher bins.}
\label{Fig08}
\end{figure}

Systematic uncertainty compositions that are fairly typical of all cross-section determinations of this work 
are shown in Fig.~\ref{Fig08}, for muon momentum (Fig.~\ref{Fig08}a) and for
pion kinetic energy (Fig.~\ref{Fig08}b).   In the lowest momentum or energy bin of
either distribution, the detector response category gives the largest fractional error; this is the result of reduced, variable 
acceptance for low-momentum muons to intercept MINOS (Fig.~\ref{Fig08}a), 
and increased uncertainty with tracking of low-energy EM showers (Fig.~\ref{Fig08}b).
At higher momentum or energy bins the detector response uncertainty diminishes and stays with a range of 3\% to 7\%.
For most bins of either distribution, finite data statistics gives a larger uncertainty than does any single systematic category.
The principal interaction cross-section model (GENIE) contributes fractional uncertainties in the range of 4\% to 10\%
for both variables, reflecting sizable uncertainties that arise with the modeling of neutrino-nucleon pion production.
Uncertainties arising from the FSI model are associated with parameters in the GENIE framework; 
the largest of which is a 50\% uncertainty associated with pion charge-exchange cross sections~\cite{Ashery-PRC-1981}.
The uncertainty propagated to a cross section measurement from the FSI-parameter uncertainties varies from 2\% to 7\%.

Uncertainties assigned to the neutrino flux are subject to constraints provided by the 
background normalization procedure, enabling this analysis to have flux-related uncertainties 
that are somewhat smaller than for most other MINERvA measurements.
The neutrino flux uncertainties in Fig.~\ref{Fig08} are roughly constant, hovering 
at or slightly below 5\% across all bins of either distribution.
Tables of measured cross-section values and of the systematic uncertainty composition 
for each bin of each measurement are provided in the Supplement to this paper~\cite{Supplement}.

\section{Muon Kinematics of CC($\pi^{0}$)}

\subsection{Muon momentum}
\begin{figure}
\begin{centering}
\includegraphics[scale=0.41]{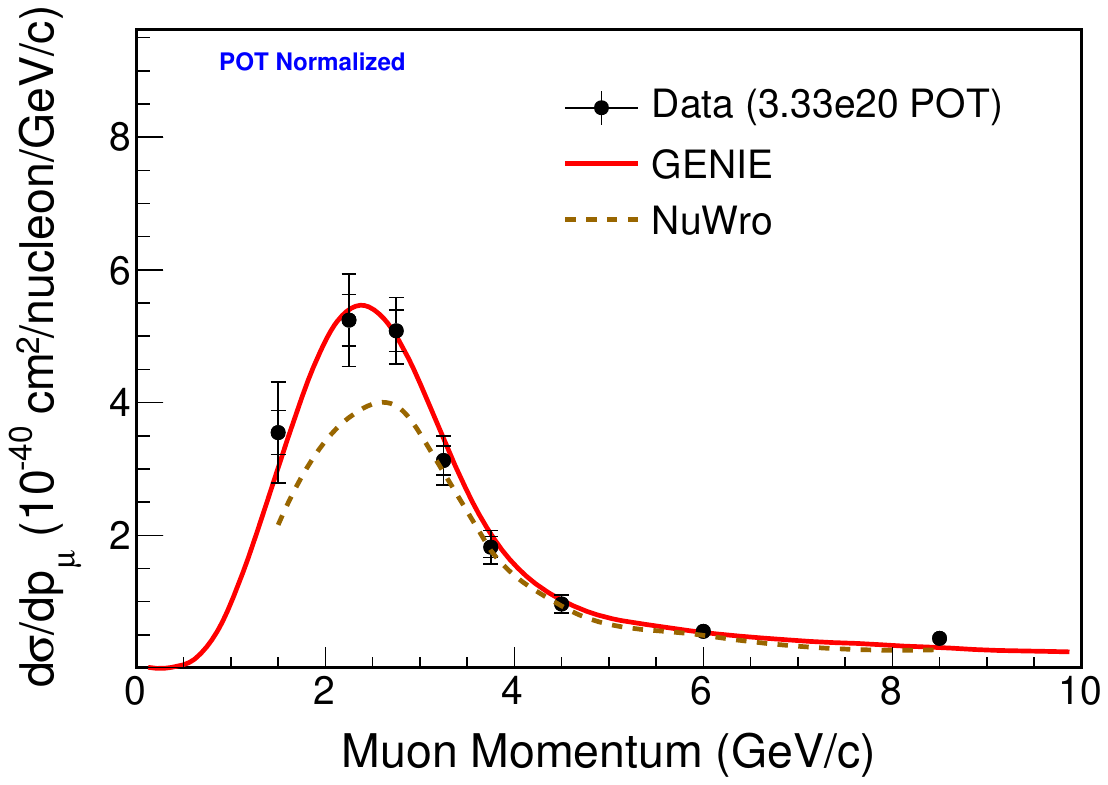}
\par\end{centering}
\caption{The flux-integrated muon-momentum differential cross section, $d\sigma/dp_{\mu}$
for muons with $\theta_{\mu} \leq 25^\circ$.  Data are shown as solid circles; 
the inner (outer) error bars correspond to statistical (total) uncertainties.   The GENIE-based simulation matches the
data distribution in shape and absolute rate to within $1\sigma$ in all bins.}
\label{Fig09}
\end{figure}

Figure~\ref{Fig09} shows the differential cross section for muon momentum, $d\sigma/dp_{\mu}$.   For this distribution and
others to follow (see Sec.~\ref{subsec:mu-p-vtx-energy}), the cross section is flux-integrated over the range 1.5\,GeV $\leq E_{\nu} <$ 20\,GeV and
the muon production angle is restricted to $\leq 25^\circ$.   The general shape of $d\sigma/dp_{\mu}$ is strongly influenced
by the $\nu$ flux spectrum; the cross section peaks between 2.0 and 2.5 GeV and falls off rapidly as $p_\mu$ increases 
from 3.0 GeV/c to beyond 6.0 GeV/c.   The GENIE-based simulation (solid-line curve) is in agreement with the data in all bins 
to within 1$\sigma$ of the statistical plus systematic error on the data distribution.    
The muon differential cross section can be compared to similar measurements
for $\numu$-CC($\pi^{+}$) and $\anumu$-CC($\pi^{0}$) obtained with \minerva-- see Fig. 7 of Ref.~\cite{Carrie-pion}.
The spectral peaks are observed to nearly coincide for all three data sets, even though the
absolute cross sections are rather different.   Differences in cross section magnitudes are to be expected, since the three
pion production channels differ in their isospin compositions and in the role played by interferences between vector current
and axial vector current contributions, the latter being constructive in the $\numu$ channels and destructive in the $\anumu$ channel.

The GENIE-based prediction in Fig.~\ref{Fig09} is less sensitive to
pion FSI than are the corresponding predictions for $\numu$-CC($\pi^{+}$) and $\anumu$-CC($\pi^{0}$).  This is 
because a full accounting of FSI alterations of pionic charge and multiplicity
predicts approximately equal feed-in versus feed-out rates for channel \eqref{signal-channel}.
The GENIE prediction in  Fig.~\ref{Fig09} is shown with FSI included; the FSI model has contributed a downwards shift in
the predicted distribution of $\sim$\,4\% in the vicinity of the peak.

In Fig.~\ref{Fig09} and in many subsequent figures, the prediction 
of the NuWro event generator~\cite{ref:NuWro} is shown with FSI effects included,
providing comparison with the GENIE-based reference simulation as well as an independent prediction for the data.  In NuWro
the baryon resonance region extends to $W < 1.6$\,GeV;  $\Delta(1232)$ production is calculated
using the Adler model~\cite{Adler-Annals-1968, Adler-PRD-1975}, with
nonresonant pion production added incoherently as a fraction of DIS, 
where DIS is formulated according to the Bodek-Yang model~\cite{Bodek-2005-PS}.
For its FSI treatment, NuWro uses the Salcedo-Oset model~\cite{Salcedo-NP-1988} 
in a hadronic cascade formalism that includes nuclear-medium corrections.
Figure~\ref{Fig09} shows that for $d\sigma/dp_{\mu}$, NuWro predicts a cross section that is lower through the peak than either the data
or the GENIE-based prediction, however it comes into agreement with the data and with GENIE at higher $p_{\mu}$ values.
Similar trends with NuWro versus GENIE predictions occur for other differential cross sections of this analysis, reflecting differences in cross-section
strengths assigned to $\Delta$ and nonresonant DIS production.

\begin{figure}
\begin{centering}
\includegraphics[scale=0.41]{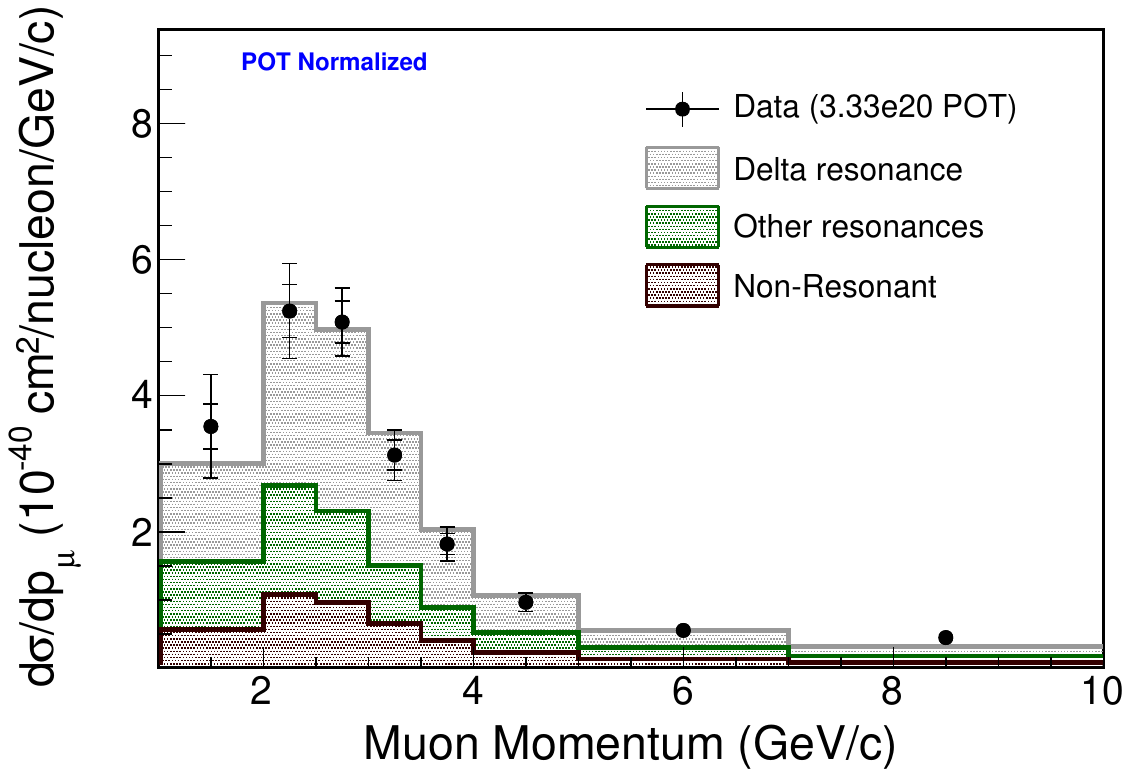}
\par\end{centering}
\caption{Composition of $d\sigma/dp_{\mu}$ in terms of three neutrino-interaction processes, 
as predicted by the simulation.  Resonance production yielding $\Delta(1232)^+\rightarrow\pi^{0} p$ is dominant (uppermost gray-shade histogram),
followed by single pion non-resonant production (bottom histogram), and production of $N^*$ states of masses above the  $\Delta^+$.}   
\label{Fig10}
\end{figure}

Events of the signal channel \eqref{signal-channel} originate in one of three general processes:  {\it i)} pion production
via the $\Delta(1232)$ resonance, {\it ii)} pion production via other baryon resonances, and {\it iii)} Non-resonant pion production. 
(As previously noted, coherent single pion production is absent.)   Figure~\ref{Fig10} shows the relative strengths of these
processes as predicted by the GENIE-based simulation.    Here, $\Delta^{+}$ production is predicted to account for  $\approx52\%$
of the rate (uppermost histogram in Fig.~\ref{Fig10});  production and decay of higher-mass $N^{*}$ resonances gives
an additional $\approx29\%$, with non-resonant single pion production accounting for the remaining 19\% of the total rate.
The relative contributions among the three processes remain nearly constant over the muon-momentum range of this $d\sigma/dp_{\mu}$
measurement.

\begin{figure}
\begin{centering}
\includegraphics[scale=0.41]{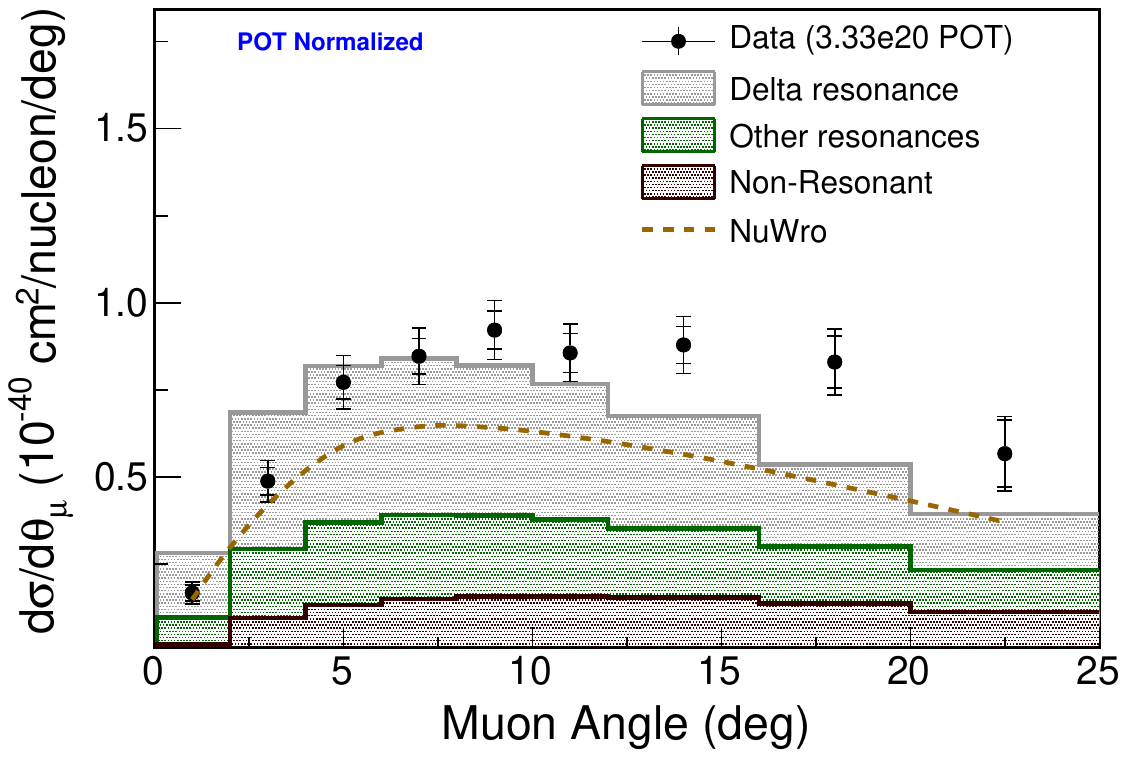}
\par\end{centering}
\caption{Differential cross section for muon production angle, $d\sigma/d\theta_{\mu}$.
The GENIE prediction (sum of histograms) exceeds the data for
$(\theta_{\mu}<5\textrm{\textdegree})$ and falls below
the data for $(\theta_{\mu}>12\textrm{\textdegree})$.  The relative contributions  
from $\Delta^+$ production, higher-mass $N^*$ production, and nonresonant $\pi^{0}$ production show little
variation over the measured range of $\theta_{\mu}$.}
\label{Fig11}
\end{figure}

\subsection{Muon production angle}
Figure~\ref{Fig11} shows the differential cross section 
as a function of polar angle, $\theta_\mu$, with respect to the 
neutrino beam.  The $\theta_\mu$ distribution peaks near $9^\circ$ and then decreases gradually at larger angles.  
The cross section is obtained from $0^\circ$ to $25^\circ$, at which point the diminishing acceptance 
for muons to reach MINOS precludes further measurement.    In contrast to the good match between
data and the reference MC observed for $d\sigma/dp_{\mu}$, significant differences are found with
the generator predictions $d\sigma/d\theta_{\mu}$.    The GENIE-based prediction overshoots the data
below $5^\circ$, while it consistently falls below the data for angles exceeding $12^\circ$.   
Milder forms of these shape disagreements can be discerned in $d\sigma/d\theta_{\mu^+}$ for $\anumu$-CC($\pi^{0}$) (data
lies below GENIE prediction at very forward angles) and in $d\sigma/d\theta_{\mu^-}$ for $\numu$-CC($\pi^{+}$) 
(data fall-off at large angles is more gradual than predicted) -- see Fig. 6 of Ref.~\cite{Carrie-pion}. 
Since muon production angle is measured with an RMS resolution 2\,mrad, the discrepancy suggests a shortfall with
the neutrino-carbon scattering model.  
These disagreements are related to data-MC disagreements observed at very low and at high $Q^2$, 
as discussed in Sec.~\ref{sec:Ev-Q2-W}.   

\begin{figure}
\begin{centering}
\includegraphics[scale=0.41]{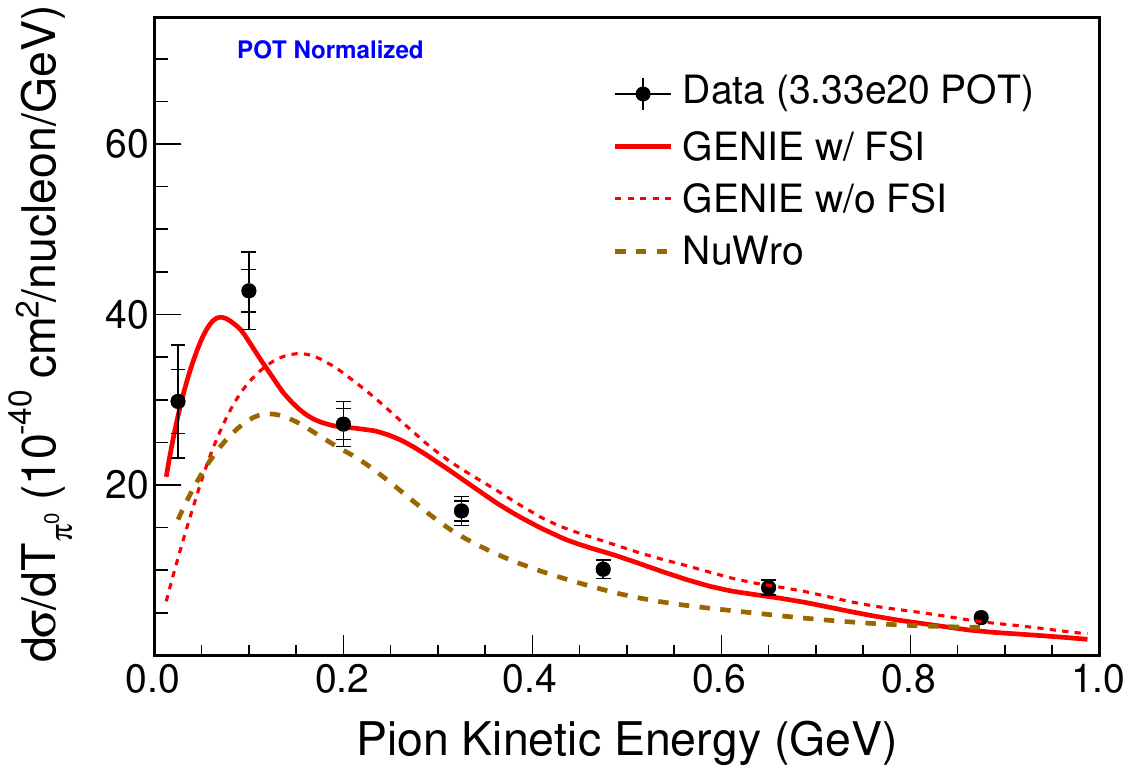}
\par\end{centering}
\caption{Pion kinetic energy differential cross section.  Predictions are shown for GENIE and NuWro; GENIE predictions
are displayed without and with pion FSI modeling (dotted-line, solid-line distributions respectively).
The data exhibits a slightly stronger enhancement at $T_{\pi^{0}} \sim 100$\,MeV than is predicted
by the simulation.   The enhancement is attributed to intranuclear charge-exchange reactions
$\pi^+ \rightarrow \pi^{0}$ fed by $\pi^+$ production from $\numu$-CC($\pi^{+}$) interactions~\cite{ Mosel-PRC91-2015}.}
\label{Fig12}
\end{figure}

\section{Pion Kinematics of CC($\pi^{0}$)}
\label{sec:Pion-Kin}
Using the three-momenta of the two EM showers in each event, 
the neutral pion momentum and kinetic energy are calculated as
$\vec{p}_{\pi^{0}}=\vec{p}_{\gamma_{1}}+\vec{p}_{\gamma_{2}}$ and $T_{\pi^{0}}=E_{\pi^{0}}-m_{\pi^{0}}$, 
where $E_{\pi^{0}}=\sqrt{\left|\vec{p}_{\pi^{0}}\right|^{2}+m_{\pi^{0}}^{2}}$.   The differential cross section for
pion kinetic energy, $d\sigma/dT_{\pi^{0}}$, is shown in Fig. \ref{Fig12}. 
The NuWro prediction lies everywhere below the data, while
GENIE generally reproduces the data normalization and shape.  
The data exhibits a rate enhancement in the region $T_{\pi^{0}} < 100$\,MeV, a trend 
predicted by the pion FSI model used by GENIE~\cite{Dytman-2011-CP}.  This effect was 
predicted by U.\,Mosel using a GiBUU simulation of \minerva data~\cite{Mosel-PRC91-2015}.  The effect is attributed
to event feed-in to channel \eqref{signal-channel} resulting from intranuclear charge-exchange involving
$\pi^+ \rightarrow \pi^{0}$; the latter processes are fueled by the large $\numu$-CC($\pi^{+}$) cross section.
It is also pointed out that the $T_{\pi^{0}}$ region around 240 MeV incurs depletion from pion absorption.

\begin{figure}
\begin{centering}
\includegraphics[scale=0.41]{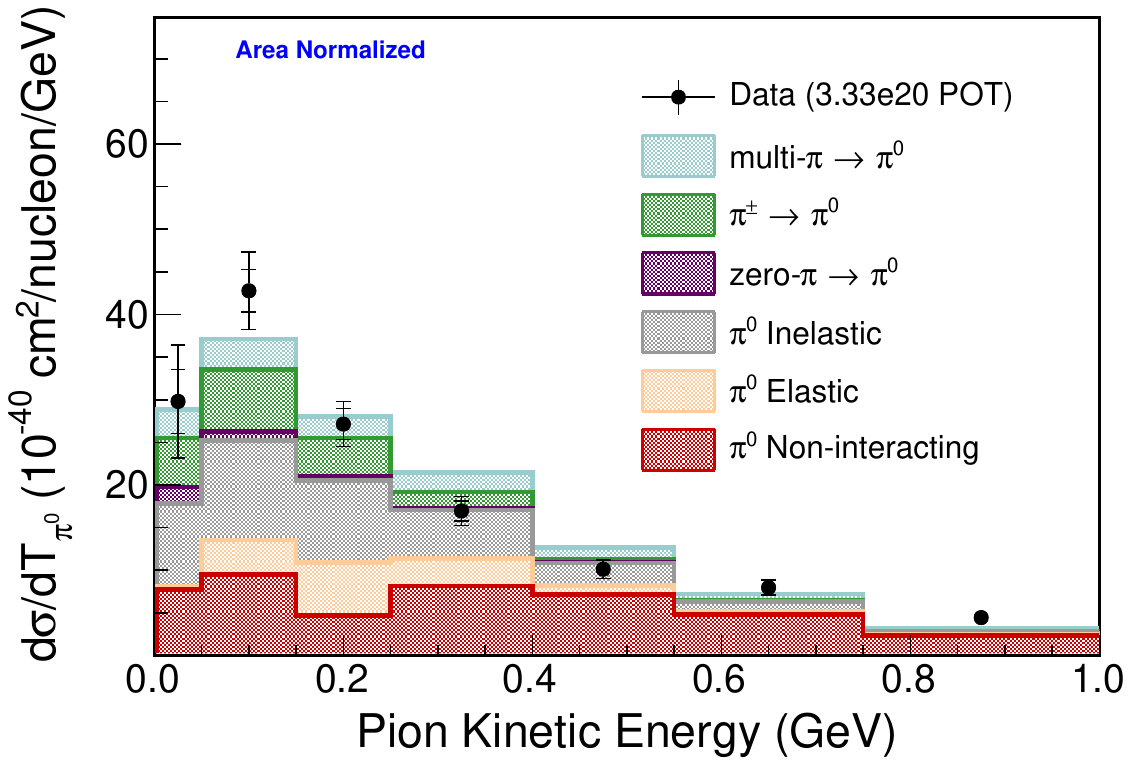}
\par\end{centering}
\caption{Composition of $d\sigma/dT_{\pi^{0}}$ in terms of 
underlying pion FSI scattering processes invoked by the GENIE model.
The FSI processes are displayed as component histograms lying above the no-scattering
case (bottom histogram).   The data and MC are shown area-normalized to each other.}
\label{Fig13}
\end{figure}

Figure~\ref{Fig13} shows the relative contributions of component pion 
scattering processes to $d\sigma/dT_{\pi^{0}}$ as invoked by the GENIE FSI model.
The data and MC distributions are shown area-normalized to each other in order to elicit
shape differences.    According to the GENIE model, $\pi^{0}$ inelastic scattering is the most probable FSI however
charge exchange $\pi^+ \rightarrow \pi^{0}$ contributes very significantly in the lowest bins of kinetic energy.
Figure~\ref{Fig13} suggests that modest tuning of pion FSI cross section may improve the data-MC agreement.

\begin{figure}
\begin{centering}
\includegraphics[scale=0.41]{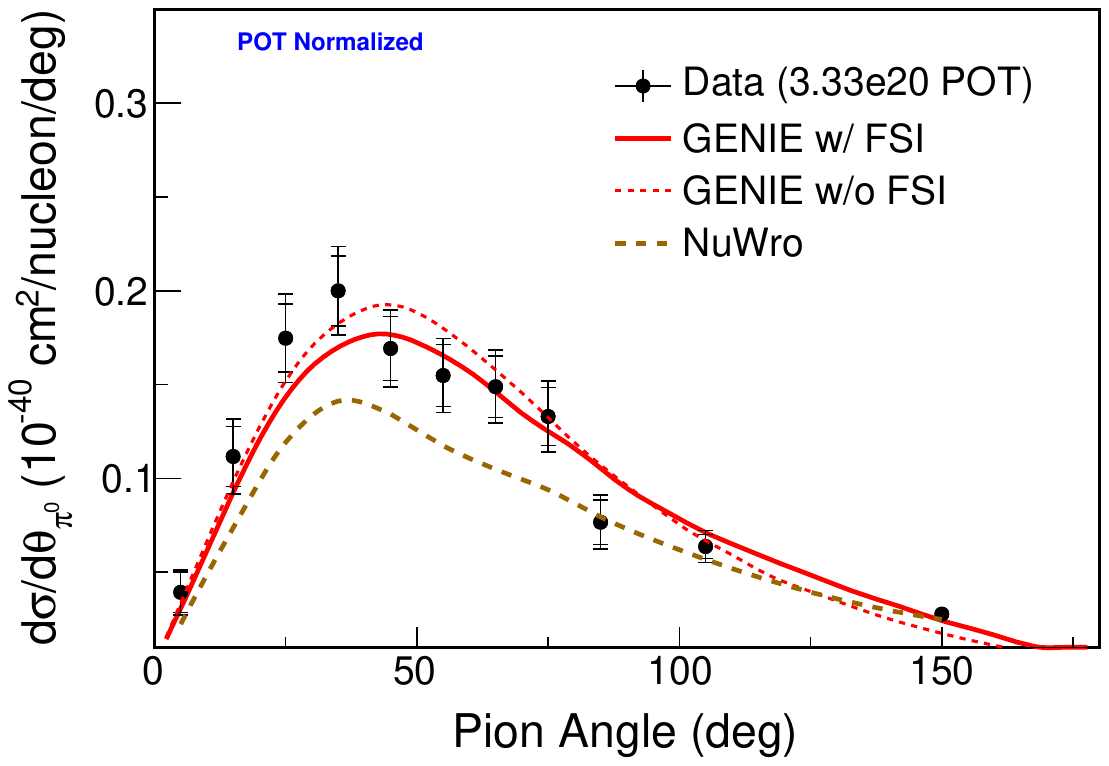}
\par\end{centering}
\caption{Differential cross section for pion angle relative to the neutrino beam direction.
The GENIE-based simulation reproduces the data distribution over the full angular range, with inclusion of pion FSI giving a
small improvement.}
\label{Fig14}
\end{figure}

The differential cross section for pion production angle with respect to the beam direction
is shown in Fig.~\ref{Fig14}.     The cross section shows most $\pi^{0}$s to be 
produced in the Lab forward hemisphere with angles around 35$^{\circ}$ being most probable. 
The GENIE-based simulation is in good agreement with the data over the entire angular range
while NuWro falls below the data in the region of the peak.   
The peak location and shape for $d\sigma/d\theta_{\pi^{0}}$ are similar to those obtained
previously for $\numu$-CC($\pi^{+}$) and $\anumu$-CC($\pi^{0}$)~\cite{Carrie-pion}; this outcome was anticipated by 
a GiBUU simulation of \minerva's low-energy exposure~\cite{Mosel-PRC91-2015}.

\section{$\sigma(E_{\nu})$,  $d\sigma$/$dQ^2$, \lowercase{and} $d\sigma/dW_{exp}$}
\label{sec:Ev-Q2-W}
\begin{figure}
\begin{centering}
\includegraphics[scale=0.41]{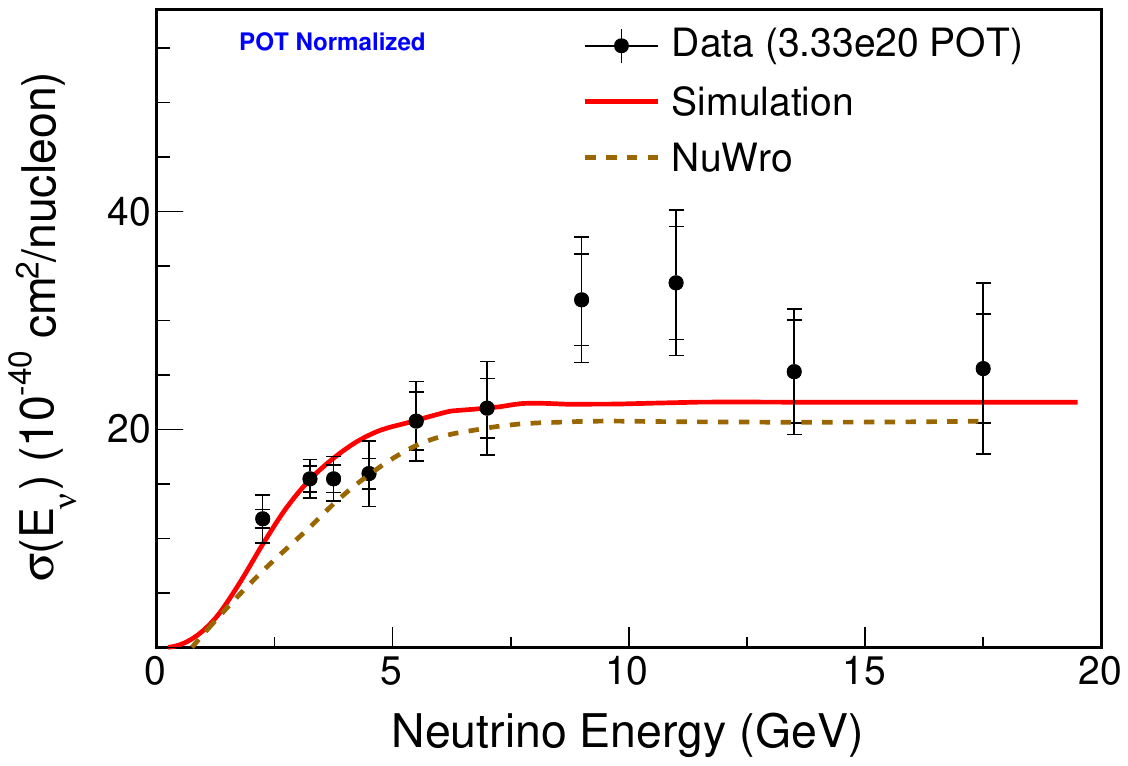}
\par\end{centering}
\caption{The cross section as a function of neutrino energy 
for the signal channel \eqref{signal-channel} 
whose definition includes $W_{exp} < 1.8$ GeV.
Data (solid circles) are shown with inner (outer) error bars 
correspond to statistical (total) uncertainties.}
\label{Fig15}
\end{figure}

Figure~\ref{Fig15} shows the channel \eqref{signal-channel} cross section as function of neutrino energy, $\sigma(E_{\nu})$, 
for events with hadronic invariant mass restricted to $W_{exp} < 1.8\,$ GeV.
The cross section rise from threshold and its value for $E_{\nu} \ge 6$ GeV 
are described on average by the reference simulation,
with NuWro predicting a cross section of similar shape and slightly lower magnitude.
According to GENIE, $\Delta$ production is the dominant process for $E_{\nu} < 3$\,GeV.   
Above 4 GeV, the contributions from $\Delta$ production, $N^*$ production, 
and nonresonant pion production become nearly independent of $E_{\nu}$ 
with relative proportions that are roughly 35:15:40~\cite{Altinok-Thesis-2017}.  
The cross section of Fig.~\ref{Fig15} can be readily compared
to the pion production cross sections reported in Ref.~\cite{Carrie-pion}.   
The measured cross sections become nearly independent of neutrino energy
around $E_{\nu} = 7$\,GeV.   At that point the relative strengths 
for $\numu$-CC($\pi^{0}$):$\numu$-CC($\pi^{+}$):$\anumu$-CC($\pi^{0}$)
in units of $10^{-40}$ cm$^2$ per nucleon, according 
to the measured cross sections, approximately follow the ratios 22:80:19.

\begin{figure}
\begin{centering}
\includegraphics[scale=0.41]{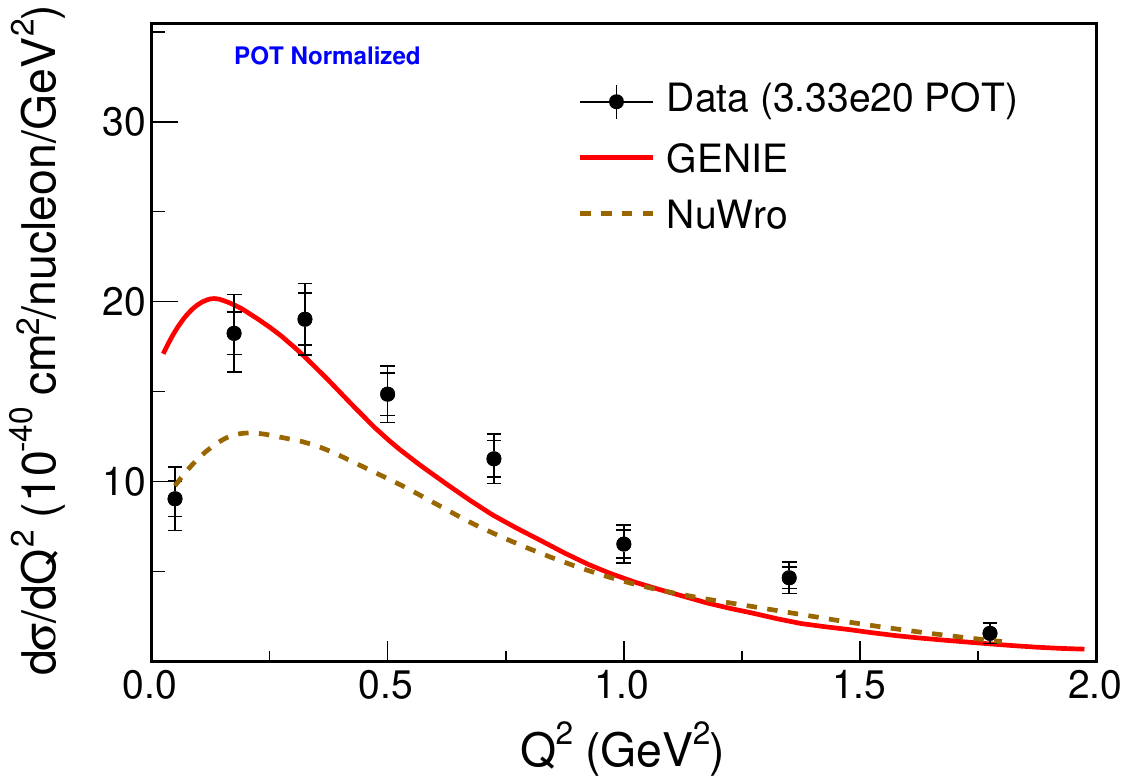}
\par\end{centering}
\caption{Differential cross section, $d\sigma/dQ^2$, for channel \eqref{signal-channel}. 
Data versus GENIE disagreements are evident for $Q^2$ near 0.0\,GeV$^2$ and for $Q^2 > 0.4$\,GeV$^2$.
These are related to the data-MC discrepancies observed at small and large muon polar angles.}
\label{Fig16}
\end{figure}

The squared four-momentum transfer from the lepton system, $Q^2$, is calculated using Eq.~\eqref{def-Q2}, which
incorporates both lepton and hadron information (via Eq.~\eqref{eq:neutrino-energy-estimate}).
Figure~\ref{Fig16} shows the differential cross section versus $Q^2$ determined by this analysis. 
The data exhibits a rate reduction at $Q^2$ below 0.2 GeV$^2$ larger than that predicted by the
GENIE-based MC.   A similar data-MC disagreement was observed 
at low $Q^2$ in the \minerva $\anumu$-CC($\pi^{0}$) sample~\cite{Carrie-pion}, and 
data suppressions at low-$Q^2$ for $\Delta$-enriched event samples have been reported
by MiniBooNE~\cite{MiniBooNE-pi0-2011, MiniBooNEPiplus:2011} and by MINOS~\cite{ref:minos-QE}.
On the other hand, the reference simulation falls below the data in the region $Q^{2}>0.4\textrm{ GeV}^{2}$.   From 
Eq.~\eqref{def-Q2} it is clear that these data-MC differences are related to those in Fig.~\ref{Fig10}
for muons produced at small and large muon polar angles.   Note that above 0.2 GeV$^2$, NuWro predicts a flatter spectrum
than does GENIE and thereby trends more similarly to the data;  however the absolute scale for $d\sigma/dQ^2$ predicted by 
NuWro is clearly too low.

For neutrino quasielastic scattering in nuclei such as carbon, 
it is well-known that Pauli blocking produces a turnover of event rate at low-$Q^2$~\cite{NOMAD-CCQE-2009}.
Additionally it is estimated on the basis of RPA and 2p2h that 
multinucleon-nucleon correlations give rise to low-$Q^2$ suppression
and high-$Q^2$ enhancement of rate in the case of 
CCQE-like scattering~\cite{Martini-mec-PRC-2009,Valencia-mec-PLB-2013, Gran-PRD-2013}.
For $\mu^{+}\Delta(1232)$ channels produced in carbon-like nuclei modeled as a Fermi gas,
the effect of Pauli blocking has been calculated in Ref.~\cite{Paschos-Pion-1}.  Pauli
suppression is shown to be confined to $Q^2 < 0.2$ GeV$^2$ and to $W < 1.4$ GeV.  
Thus Pauli blocking of $\Delta$ and $N^*$ states, which is not included in the reference simulation,
plausibly accounts for a modest portion of the low $Q^2$ suppression exhibited by the data in Fig.~\ref{Fig16}.
It is possible that NN-correlation effects of the kind targeted 
by RPA calculations may also be present in neutrino-nucleus baryon resonance
production, however this has yet to be demonstrated with a calculation.  
If the calculation of Ref.~\cite{Gran-PRD-2013}, which is for quasielastic scattering (and not baryon resonance production),
is indicative of the strength and $Q^2$ dependence of RPA distortion, then RPA in the absence of 2p2h
is conceivably capable, in conjunction with Pauli blocking, of generating the data-MC disagreement shown
in Fig.~\ref{Fig16}.   Admittedly, the latter scenario is speculative and it may be at odds with a recent theoretical treatment
that shows RPA correlations to have reduced effect in calculations based on a realistic nuclear ground state~\cite{Pandey-PRC-2016}.
Another possibility, recognized for many years~\cite{Paschos-Pion-1}, is that the normalization or functional form of the
axial-vector form factors of neutrino-nucleus scattering may need modification.   Unfortunately there is a dearth of theoretical guidance on
how to approach this.

\begin{figure}
\begin{centering}
\includegraphics[scale=0.42]{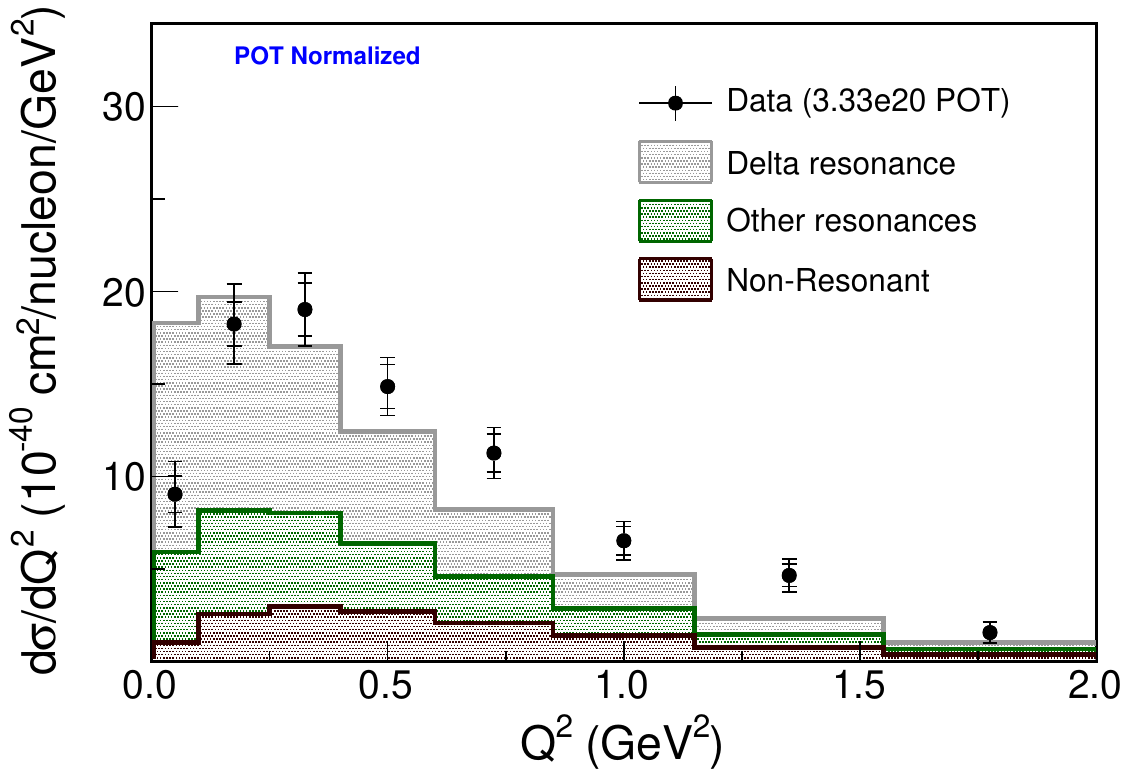}
\par\end{centering}
\caption{Reaction composition for $d\sigma/dQ^{2}$
for the sample  $\nu_{\mu}\textrm{-CC}(\pi^{0})$.  Production of $\Delta^{+}(1232)$ (uppermost histogram) dominates
the $Q^2$ region below 0.4\,GeV$^2$.   Production of higher-mass $N^*$ states and of nonresonant $\pi^{0}$'s becomes
increasingly important at higher $Q^2$.}
\label{Fig17}
\end{figure}

Figure~\ref{Fig17} displays the reaction composition of $d\sigma/dQ^{2}$ as predicted by the 
GENIE-based MC.  Baryon-resonance production, 
especially $\Delta(1232)$ production, is expected to dominate at low $Q^2$.   
However as $Q^2$ is increased, nonresonant pion production gradually
overtakes the baryon resonance contributions to become the dominant single process for $Q^2$ beyond 1.0 GeV$^2$.

\begin{figure}
\begin{centering}
\includegraphics[scale=0.42]{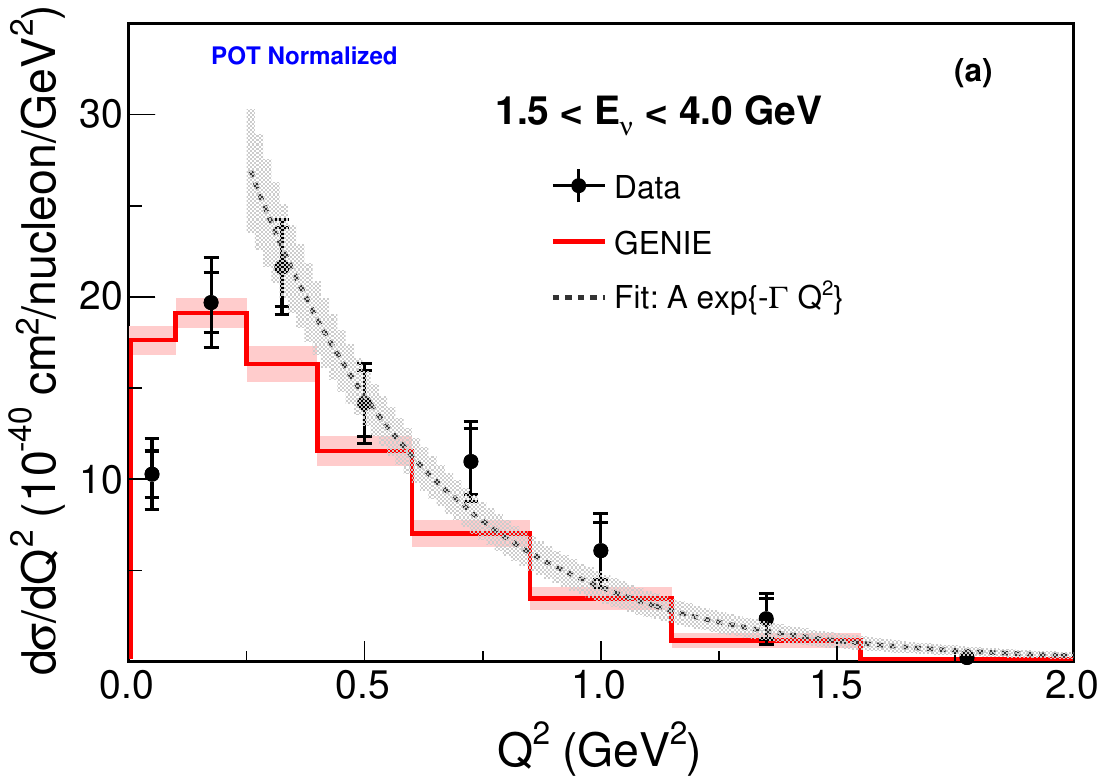}
\includegraphics[scale=0.42]{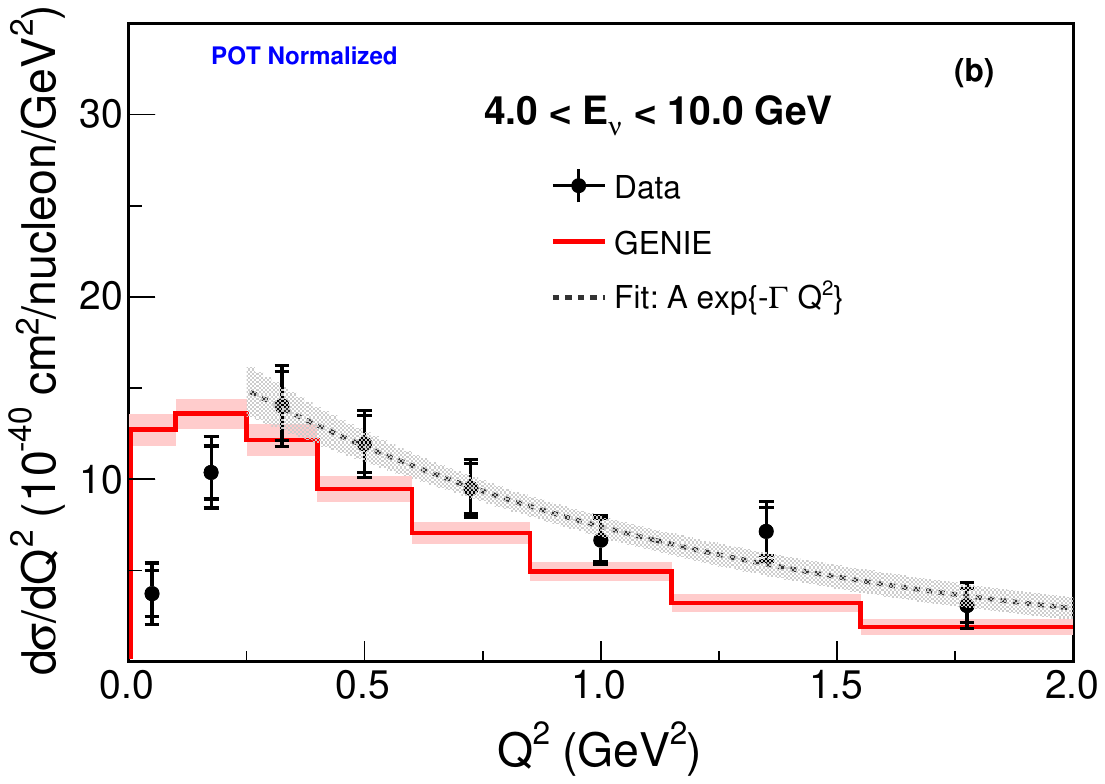}
\par\end{centering}
\caption{The $d\sigma/dQ^2$ distributions for signal events separated into low (a) and high (b) ranges of $E_{\nu}$.  
For events with $0.25 < Q^2 < 2.0$\,GeV$^2$, the falloff with increasing $Q^2$ is approximately exponential and exhibits a steeper slope for the lower-$E_{\nu}$ sample.}
\label{Fig18}
\end{figure}

The relative contributions from $\Delta(1232)$ production, higher-mass $N^*$ production, 
and nonresonant pion production may be expected to change with neutrino energy, 
with $N^*$ and nonresonant contributions becoming more important with increasing $E_{\nu}$.    
From a different perspective, namely that of weak-interaction hadronic currents, 
$d\sigma/dQ^2$ can be considered as the sum of vector and axial-vector  terms, 
plus the VA interference term which is constructive for the case of CC neutrino scattering.   
For incident $E_{\nu}$ of one to few GeV, VA interference may exceed the vector contribution 
however its presence diminishes very rapidly at higher $E_{\nu}$~\cite{Lalakulich-2005}.  
And of course, the phase space available to the final state grows with $E_{\nu}$.   
The interplay of these effects can give rise to an $E_{\nu}$ dependence for $d\sigma/dQ^2$.   
To elicit such a dependence, determinations of $d\sigma/dQ^{2}$ have 
been carried out for two different energy ranges, namely $1.5 < E_{\nu}<4\textrm{ GeV}$ and $4 < E_{\nu}<10\textrm{ GeV}$.   
Figure~\ref{Fig18} shows $d\sigma/dQ^{2}$ for signal distributions separated into the two energy ranges
according to reconstructed $E_{\nu}$.
The difference in the rate of falloff with increasing $Q^2$ between the two samples is evident.

For the purpose of providing a phenomenological characterization of the data trend, the 
shapes of $d\sigma/dQ^{2}$ for $0.25 < Q^2 < 2.0$\,GeV$^2$ are fitted to an exponential decay function, 
and the slope parameters, $\Gamma \equiv$ 1/$Q_{0}^2$, are obtained separately
for the low-$E_{\nu}$ and high-$E_{\nu}$ ranges.
The slope values from the fits are
\begin{equation}
\label{slope:low-Ev}
\begin{split}
&\Gamma_{\text{low}\,E_{\nu}} ~=  2.55\,\pm 0.26~\text{GeV}^{-2}~~~~\text{and} \\ 
&\Gamma_{\text{high}\,E_{\nu}} =  0.93\,\pm 0.21~\text{GeV}^{-2}.
\end{split}
\end{equation}
Thus the slope of $d\sigma/dQ^{2}$ flattens as $E_{\nu}$ is raised from few GeV to multi-GeV values.  
This trend is roughly reproduced by the GENIE-based reference simulation.

\begin{figure}
\begin{centering}
\includegraphics[scale=0.42]{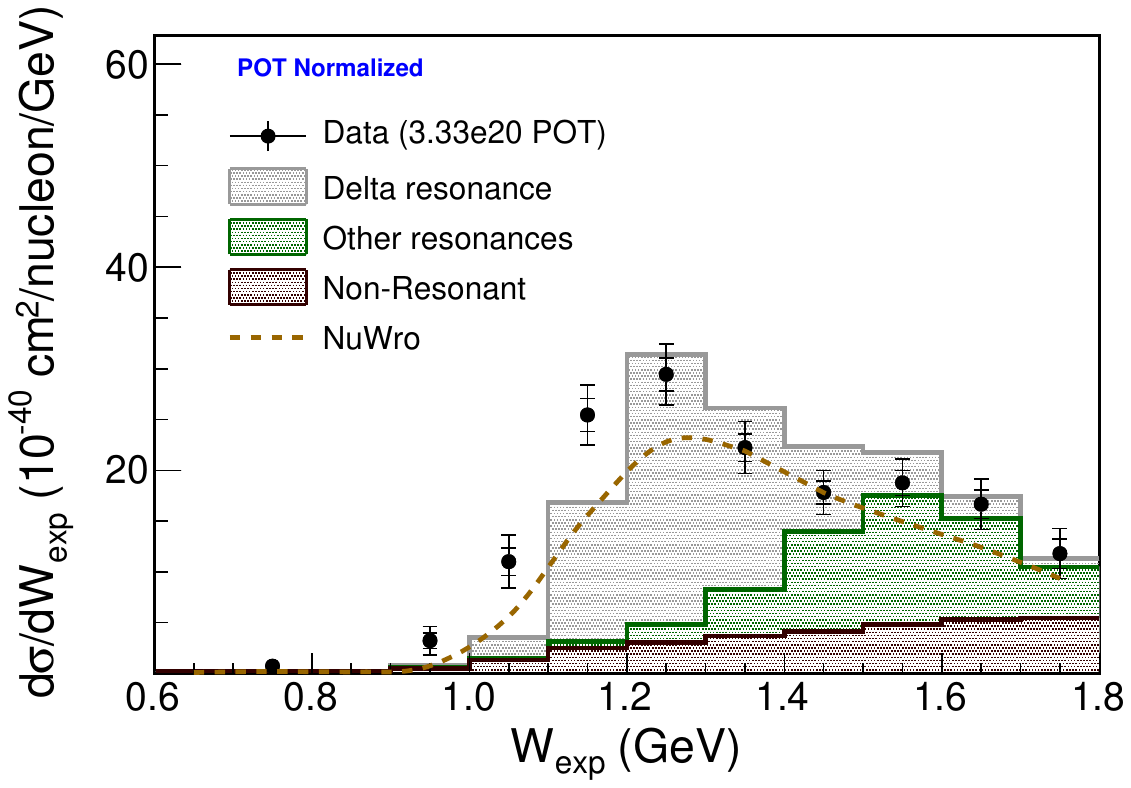}
\par\end{centering}
\caption{Distribution of $W_{exp}$, the hadronic mass estimated 
using Eqs.~\eqref{def-Q2} and \eqref{def-W}.   The data 
peaks in the vicinity of the $\Delta(1232)^{+}$ resonance.  The reference simulation
roughly matches the data rate, but the predicted shape is shifted
towards higher $W_{exp}$ values relative to the data.}
\label{Fig19}
\end{figure}

The event distribution of reconstructed hadronic mass, $W_{exp}$, in the data and as predicted using GENIE or NuWro,  
is shown in Fig.~\ref{Fig19}.     The reference simulation and NuWro as well fail to match the
data shape in regions below and above the  broad spectral peak near the $\Delta(1232)^{+}$ resonance.  
The disagreement is of an unusual kind;
the predictions are shifted towards higher $W_{exp}$ relative to the data.   The average displacement is estimated by imposing an
overall shift on the prediction and calculating the $\chi^{2}$ with respect to data for each trial.    A shift of 20 MeV gives the minimum
$\chi^2$, improving the $\chi^{2}/d.o.f.$ from 6.9 to 1.8.  
As indicated by Eq.~\eqref{def-W} from which $W_{exp}$ is calculated, a portion of this shift may reflect 
the data-MC disagreement in $Q^2$ exhibited by Figs.~\ref{Fig16} and \ref{Fig17}.  
According to simulation trials, the $Q^2$ offsets can introduce shifts of up to 10 MeV in the distribution of $W_{exp}$.  
Equation~\eqref{def-W} is founded on the unrealistic assumption that target nucleons are struck while at rest, 
and this makes the MC predictions sensitive to shortfalls in modeling of interactions on nuclei.  
The differences in modeling approaches used by the GENIE and NuWro neutrino generators
provide some perspectives.   Alteration of Fermi-motion modeling and inclusion of in-medium modification 
of the $\Delta(232)$ resonance~\cite{Agababyan-2013}, 
may generate offsets of $\sim$\,5 MeV.   A shift of similar magnitude may also arise from constructive interference 
between the $\Delta$ and nonresonant pion production amplitudes [16].   
The latter mechanism is not treated by either generator.

\section{Subsamples with $p \pi^{0}$ systems}
To study the production of high-mass $N^*$ states and especially of the $\Delta(1232)^+$,
it is useful to define two subsamples of the analysis signal sample.   
Events of these subsamples are required to have a leading proton in the final state 
with kinetic energy $T_{p} > 100$\,MeV.    For the first subsample, 
designated hereafter as the $p\pi^{0}$ sample, the signal definition requirement 
$W_{exp}<1.8$\,GeV is left unchanged.   For the second subsample, 
designated as the $\Delta$ enriched-subsample the $W_{exp}$ range is
restricted to $W_{exp}<1.4$\,GeV.   The added requirements amount 
to a new signal definition for each subsample,  consequently the procedures 
used for the signal sample of channel \eqref{signal-channel}, namely background fitting, 
background subtraction, unfolding, and efficiency correction, have been carried out anew for each subsample.   For
either subsample, the invariant mass, $M_{p\pi^{0}}$, can be 
calculated from the reconstructed $\pi^{0}$ and the leading proton.
The resolution (RMS width) for $M_{p\pi^{0}}$ is 0.10 GeV.

\begin{figure}
\begin{centering}
\includegraphics[scale=0.43]{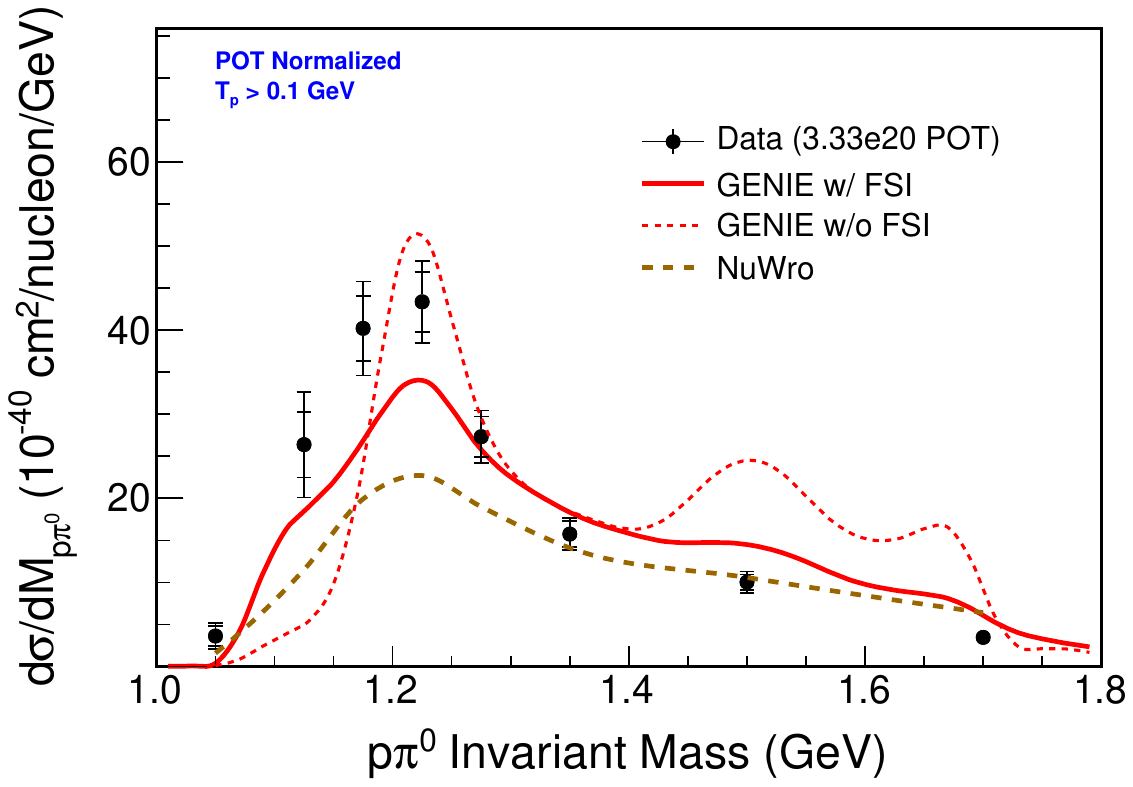}
\par\end{centering}
\caption{Cross section versus $M_{p\pi^{0}}$ for the $p\pi^{0}$ sample, requiring $T_{p} > 100$\,MeV with $W_{exp}<1.8$\,GeV.
Curves predicted by the reference simulation show that hadronic FSI tends to broaden and mute baryon-resonance structures.
In the $\Delta(1232)^+$ region however, the data exhibits a resonance shape that is more pronounced than that predicted by 
either the GENIE or NuWro generators.}
\label{Fig20}
\end{figure}

The $p\pi^{0}$ sample prior to background subtraction consists of 3316 events 
with an estimated signal purity of 54\%.    Figure~\ref{Fig20} shows the differential cross section 
in $M_{p\pi^{0}}$ obtained with this sample.  The GENIE-based simulation 
is seen to be strongly affected by the FSI model.   The presence of multiple 
baryon-resonance states above 1.4 GeV is clearly discernible in the simulation 
without FSI (short-dash curve), but the FSI effectively washes away the $N^*$ peaks 
to give a distribution (solid curve) that more nearly describes the data.   Both GENIE and NuWro 
however appear to underestimate the amount of $\Delta^+$ production that is indicated by the data.

\begin{figure}
\begin{centering}
\includegraphics[scale=0.41]{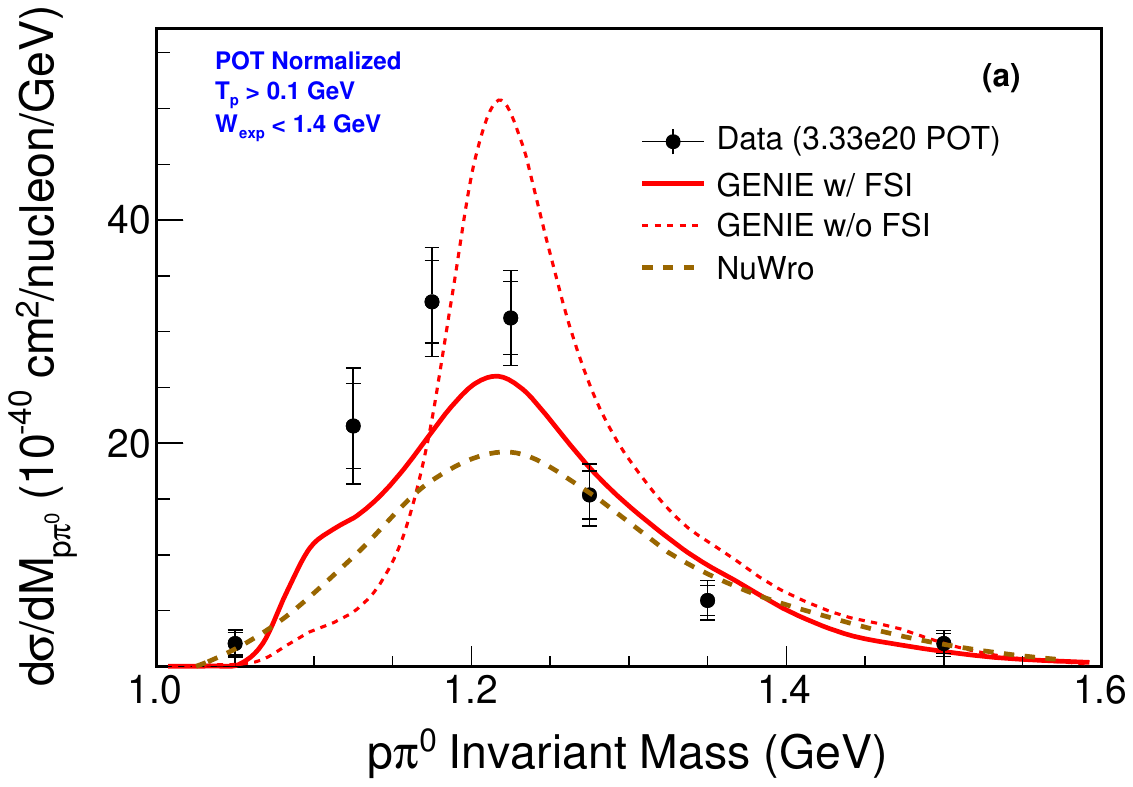}
\includegraphics[scale=0.41]{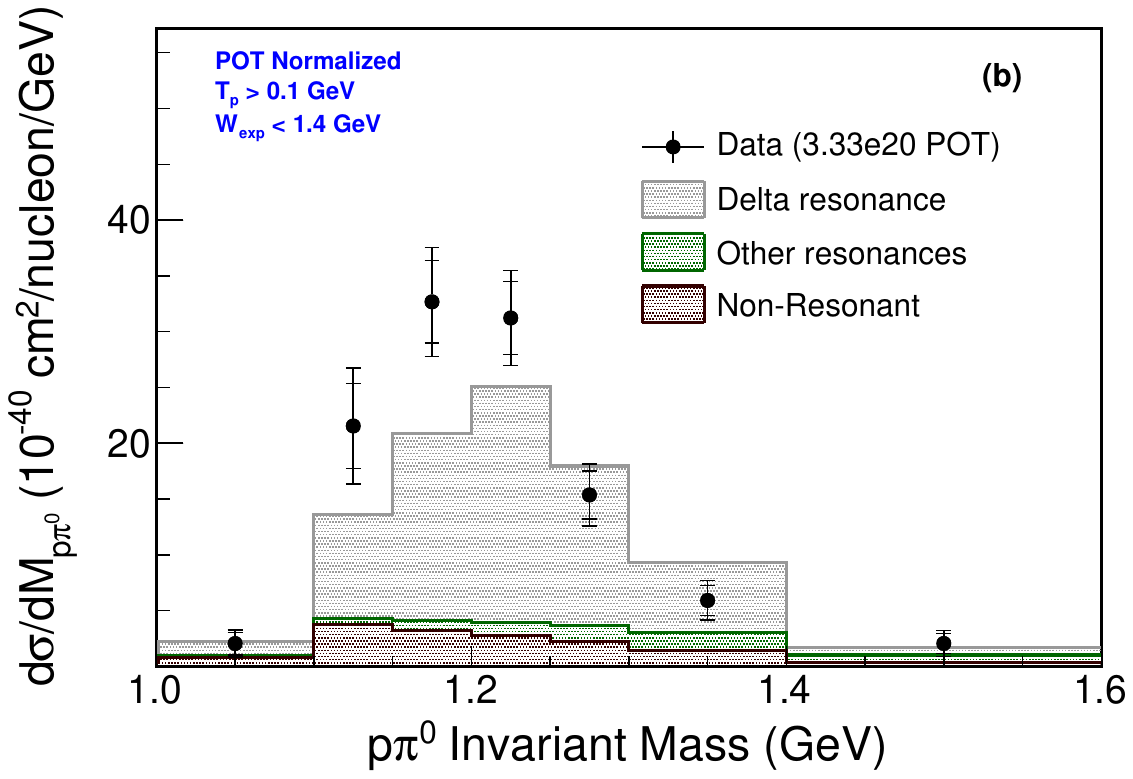}
\par\end{centering}
\caption{Proton-pion invariant mass for the $\Delta$-enriched subsample with $W_{exp} < $1.4\,GeV and a reconstructed proton
in each event.   GENIE with FSI (solid curve in (a)) is broader and falls below 
the data near the $\Delta^{+}$ peak.   The $M_{p\pi^{0}}$ distribution is dominated by produced
$\Delta^{+}$ states (uppermost histogram in (b)).  For both GENIE and NuWro, the predicted shape is shifted 
towards higher $M_{p\pi}$ relative to the data.}
\label{Fig21}
\end{figure}

Study of produced $\Delta^{+}$ states in isolation from the higher-mass resonances is 
carried out using the $\Delta$-enriched subsample.   The subsample contains 1522 events and
has a 46\% signal purity.   Figure~\ref{Fig21}a shows $d\sigma/dM_{p\pi^{0}}$ for the 
subsample, comparing the data (solid points) to GENIE and NuWro predictions.
The GENIE prediction with FSI included (solid curve) falls below the data 
in the $\Delta$ region $1.15 < M_{p\pi}<1.30$ GeV.   However the predicted distribution 
is broader which, with respect to total rate, partially compensates for its rate shortfall 
at the resonance peak.  NuWro predicts a smaller cross section at the $\Delta$ peak than does GENIE, but gives
a similar cross section at higher invariant mass values.

Figure~\ref{Fig21}b shows the subsample in terms of three main interaction categories.  
According to GENIE, the subsample is composed of 74\% $\Delta^{+}$,~10\% higher-mass 
$N^*$ production, and 16\% nonresonant pion production.   The $M_{p\pi}$ prediction relative 
to the data exhibits the same trend towards higher hadronic invariant mass values as observed in Fig.~\ref{Fig19}.

\subsection{Polarization of the p$\pi^{0}$ system}

Bubble chamber investigations of neutrino-induced $\Delta^{++}$ production in the few-GeV region of $E_{\nu}$   
found the resonant state to be polarized, as evidenced by non-isotropic $\pi^+$ angular distributions of $\Delta^{++}$ decays
measured in the  $p\pi^+$ rest frame~\cite{Schreiner-VonHippel-1973, Radecky-1982, Allasia-1990}.   
In general, the decay angular distributions of produced $\Delta$ states are sensitive to production mechanisms and to interferences 
among different partial wave amplitudes, hence measurements for other $\Delta$ charge states can provide
new tests of resonance production models~\cite{Rein-Sehgal, Rein-Z-1987, Allen-NP-1986}.
In this work the produced $p\pi^{0}$ systems of the $\Delta$-enriched subsample are examined for
deviations from isotropic decay that may reflect polarization effects.
The decay angles for $\pi^{0}$ mesons emitted in the $p\pi^{0}$ rest frame are calculated using
formalism suited to this measurement~\cite{Adler-Annals-1968, Schreiner-VonHippel-1973}.
The coordinate system used in the $p\pi^{0}$ rest frame is established event-by-event as follows: 
The four-momenta of the neutrino, muon, pion
and proton are Lorentz-boosted into the $p\pi$ rest frame.  In that frame, the axes of a 
right-handed coordinate system are defined with the
$z\textrm{-axis}$ along the momentum transfer direction $(\vec{p}_{\nu}-\vec{p}_{\mu})$,
the $y\textrm{-axis}$ along the production plane normal $(\vec{p}_{\nu}\times\vec{p}_{\mu})$,
and $x\textrm{-axis}$ along the direction of the cross-product 
$(\hat{y}\times\hat{z})$.  The zenith angle $\theta$ is the
angle between the pion momentum, $\vec{p}_{\pi}$, and
the $z\textrm{-axis}$. The azimuthal angle $\phi$ is the angle between the
projection of $\vec{p}_{\pi}$ on the $x-y\textrm{ plane}$ and $x\textrm{-axis}$.

\begin{figure}
\begin{centering}
\includegraphics[scale=0.37]{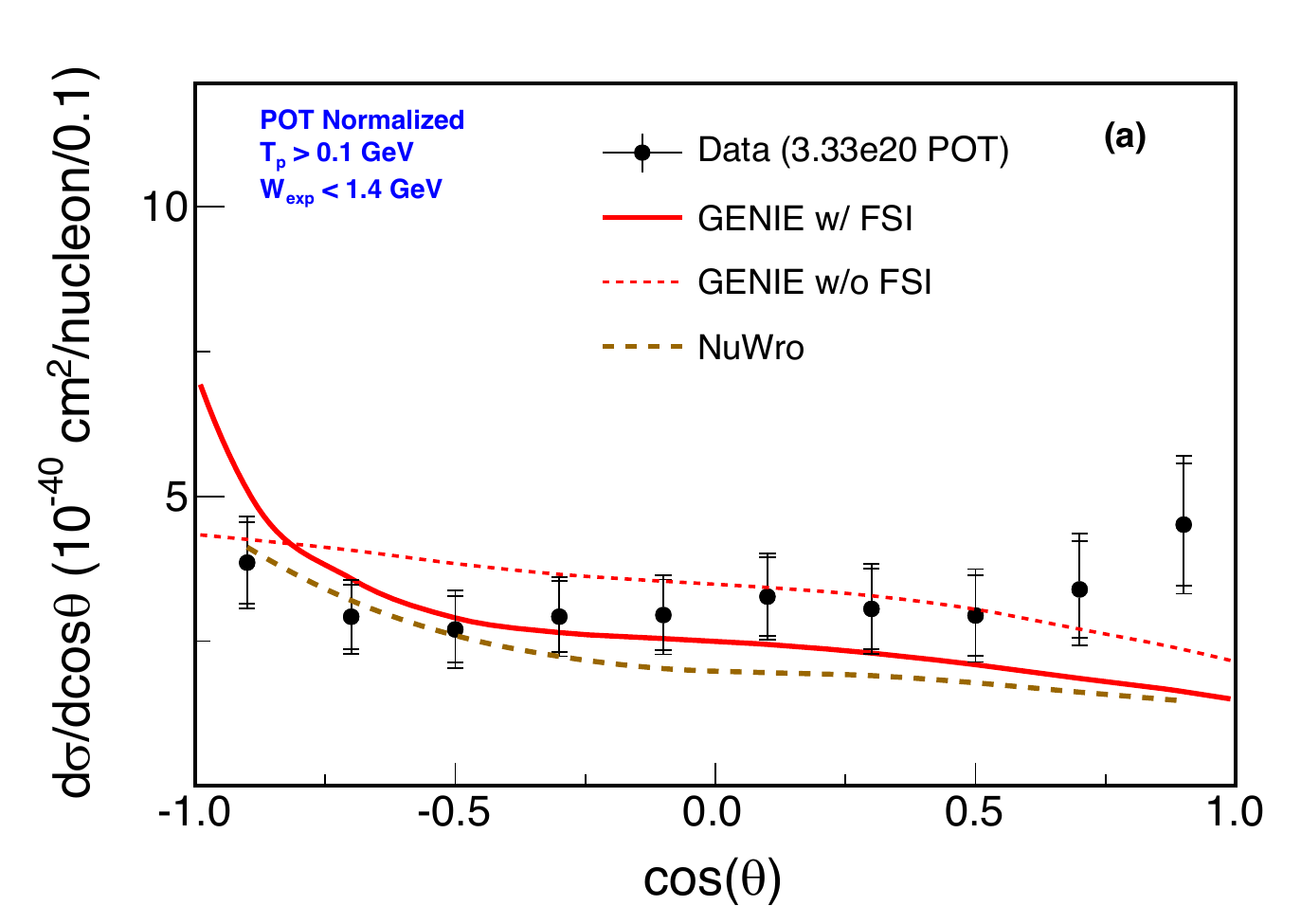}
\includegraphics[scale=0.37]{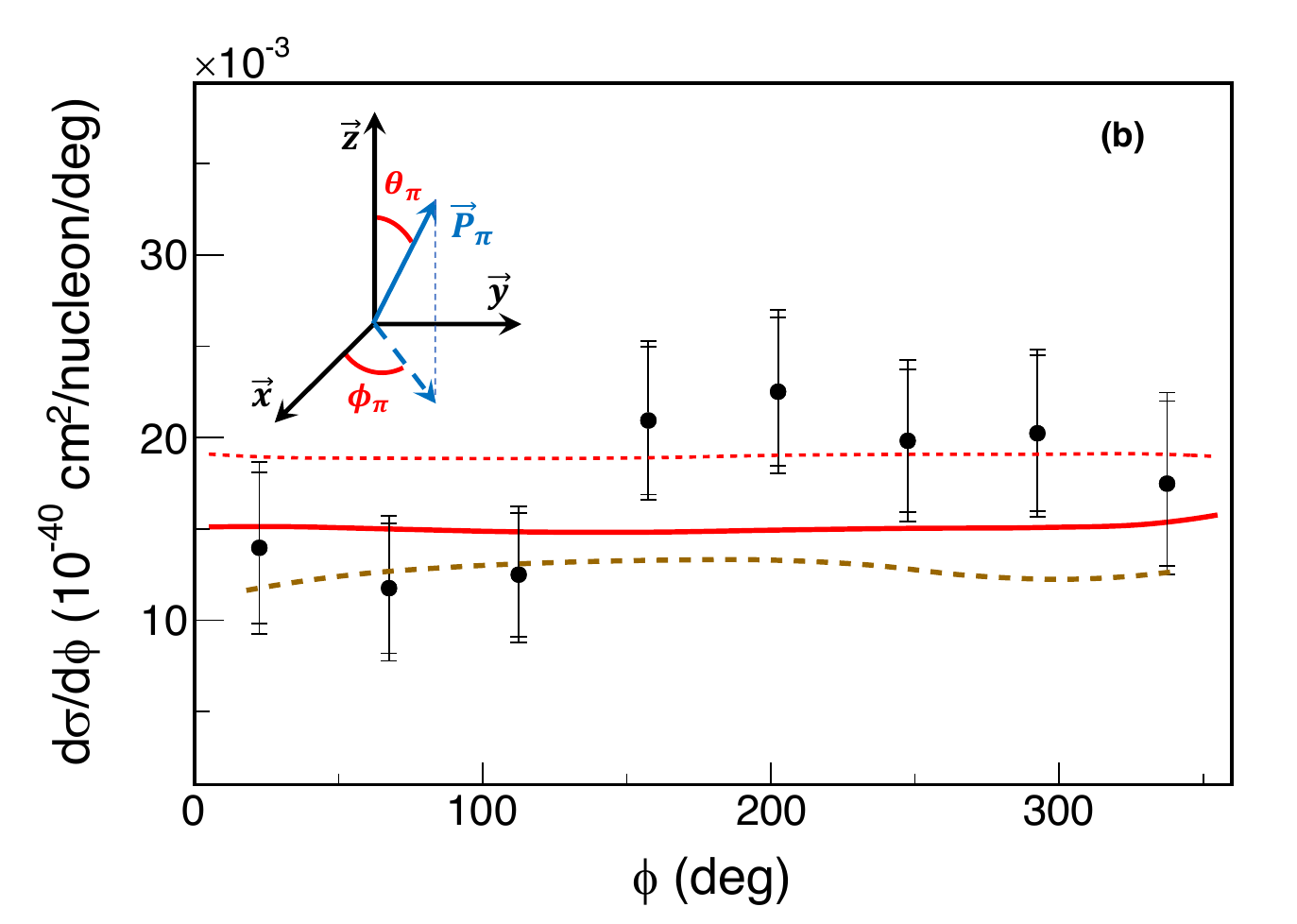}
\par\end{centering}
\caption{Angular distributions in zenith-angle cosine (a) and azimuthal angle $\phi$ (b)
for pion emission in the $p\pi^{0}$ rest frame, from the  data (solid circles), the GENIE-based simulation (solid lines), and
the prediction of NuWro (dashed lines).   
The generator models assume $\Delta^+$ resonances to decay isotropically in their rest frames and they roughly 
describe the data, however mild discrepancies are discernible in the forward hemisphere ($\cos(\theta) > 0.0$) and
with respect to left-right (a)symmetry about $\phi = 180^{\circ}$.}
\label{Fig22} 
\end{figure}

The distributions of zenith-angle cosine, $\cos(\theta)$, for the
data and as predicted by the GENIE-based simulation and by NuWro, and are shown in Fig.~\ref{Fig22}a.
For the predictions by the neutrino generators, the two-body decays of baryon resonances, including $\Delta^+ \rightarrow p + \pi^{0}$, are 
generated isotropically in their rest frames.   (Exception is made in MINERvA's GENIE implementation for $\Delta^{++}$ decays, where
polarization (for $\cos(\theta)$) at 50\% of the strength prescribed by the Rein-Sehgal model is used~\cite{Brandon-pion}.)
According to the GENIE, $p\pi^{0}$ systems that do not experience FSI would distribute fairly isotropically.  
In the presence of FSI however a peak develops in the backwards hemisphere 
for $\cos(\theta) < -0.5$ as shown by the dotted-line and solid-line curves in Fig.~\ref{Fig22}a.
A similar trend is predicted by NuWro, and indeed the data shows a mild upswing as $\cos(\theta)$ approaches -1.0. 
This peaking in the backwards direction is a pion FSI effect.   As can be seen in Fig.~\ref{Fig12}, FSI enhances the low-momentum
component of the pion spectrum.   When slow pions are boosted to the $p\piz$ rest frame (the boost direction in the Lab being roughly aligned
with the direction of proton momenta), they are projected into the backwards hemisphere where they are anti-aligned with the protons and
with the boost direction which defines the z-axis.

In the forward hemisphere the data also shows a mild increase in rate at very forward directions; moreover the data lies above the GENIE
and NuWro predictions throughout the foward hemisphere.   While the absolute event rates in the very forward hemisphere are sensitive
to the value set as the proton threshold requirement ($T_{p} > 100$\,MeV), 
the tendency for the data to exceed the predictions throughout the forward hemisphere 
is invariant to changes in this threshold setting.    The overall trend in data versus prediction based upon isotropic decay
is suggestive of weak polarization excitation of density matrix elements associated 
with a $Y_{2}^{\pm 1}(\theta, \phi)$ angular dependence~\cite{Rein-Sehgal}.

Figure~\ref{Fig22}b compares the data and simulation prediction for the distribution
of azimuthal angle $\phi$.  Here, FSI is predicted to give an overall reduction in rate (as is predicted for $\cos(\theta)$), however
neither FSI nor details of the subsample selection introduce any angular distortions; the shape of the $\phi$ distribution is predicted
to be flat -- consistent with isotropy -- by both event generators.   To zeroth order the data is also isotropic, however there are deviations
at the $< 2$\,$\sigma$ level that suggest a left-right asymmetry relative to the X-Z plane, with $0^{\circ} < \phi <180^{\circ}$ defining
the ``Right" side of the plane.    

In summary, the data -- when compared to simulations based upon isotropic decays 
for produced $\Delta^+$ states -- show modest deviations from isotropy with respect to both $\cos(\theta)$ and $\phi$.   
These anisotropies can be quantified using two test statistics, namely a forward/backward, data-vs-MC, ratio-of-ratios
to express asymmetry about the plane at $\cos(\theta) = 0.0$, and a left/right, data-vs-MC, ratio-of-ratios to express asymmetry 
with respect to $\phi = 180^{\circ}$.  These ratios are akin to the asymmetry parameters 
reported by the ANL bubble chamber experiment~\cite{Radecky-1982}.   
Let $N_ {F(B)}$ and $N_{L(R)}$ designate number of events in forward (backward) and left (right) hemispheres
respectively, and let $R_{F/B}$ and $R_{L/R}$ designate $N_{F}/N_{B}$ and $N_{L}/N_{R}$ respectively; then

\begin{equation}
\label{ratio-of-ratios}
\frac{R_{F/B}^{(data)}}{R_{F/B}^{\text{(GENIE)}}}  = 1.99 \pm 0.41,  ~
\frac{R_{L/R}^{(data)}}{R_{L/R}^{\text{(GENIE)}}}  = 1.36 \pm 0.27.
\end{equation}

\noindent
Here, the prediction of the GENIE-based simulation is taken to represent isotropy, and anisotropy is 
gauged in terms of a ratio-of-ratios deviation from unity.
Thus the data exhibits a $\sim 2\,\sigma$ anisotropy that favors pion emission into the forward hemisphere, and a 
$\sim 1\,\sigma$ anisotropy that favors emission into the left hemisphere with $180^{\circ} < \phi < 360^{\circ}$.   
Comparing to the asymmetries observed with $\Delta^{++}$
states produced in events of the ANL bubble chamber experiment (with $0.5 < E_{\nu} < 6$\,GeV)~\cite{Radecky-1982}, the 
anisotropies in $\cos(\theta)$ and $\phi$ are of comparable magnitudes and of opposite and same sign, respectively.

\section{Conclusions}
\label{sec:Conclusions}

The measurements of this work provide a multifaceted view of semi-exclusive $\numu$-CC($\pi^{0}$) 
scattering on carbon.   Differential cross sections are presented for muon variables $p_{\mu}$ 
and $\theta_{\mu}$ and for $T_{\pi}$ and $\theta_{\pi}$ of the final-state $\pi^{0}$.   
The per event $E_{\nu}$ is estimated from muon kinematics 
plus a sum over calorimetric measures of final-state hadronic energy, and cross 
sections are thereby determined as functions of $E_{\nu}$, $Q^{2}$, and $W_{exp}$. 
From the signal sample, events having a proton above the reconstruction threshold ($T_{p} > 100$\,MeV) 
are selected.  Differential cross sections in proton-$\pi^{0}$ invariant mass are reported for
events with $W_{exp} < 1.8$\,GeV, and also for events with the further selection $W_{exp} < 1.4$\,GeV.
Using events of the latter subsample, the angular emission 
of pions in the $p\pi^{0}$ rest frame is examined for evidence of
polarization, and differential cross sections for zenith-angle $\cos(\theta)$
and for azimuthal angle $\phi$ are obtained.

The $\nu_{\mu}\textrm{-CC}(\pi^{0})$ cross-section
distributions are compared to a modified GENIE simulation 
used by other recent MINERvA studies and to the predictions of NuWro.  They are also 
compared to previous MINERvA measurements of the single-pion
production channels $\nu_{\mu}\textrm{-CC}(\pi^{+})$ and 
$\bar{\nu}_{\mu}\textrm{-CC}(\pi^{0})$~\cite{Brandon-pion, Trung-pion, Carrie-pion}.
Summary tables for the cross-section measurements that may facilitate data comparisons and 
phenomenological study are available in the Supplement~\cite{Supplement}.

These measurements promote the development of neutrino interaction models that will, one day, encompass
the physics that underwrites CC single pion production from nuclear targets.   Since the CC($\pi^{0}$) channel occurs at
significant rate in long baseline neutrino detectors, and since the differential cross sections obtained
span the working $E_{\nu}$ range used by all accelerator-based $\nu$ oscillation experiments, 
the results reported here will enable continued improvement in precisions achievable with oscillation measurements.

\section*{Acknowledgments}
 This work was supported by the Fermi National Accelerator Laboratory under US Department of Energy contract
No. DE-AC02-07CH11359 which included the MINERvA construction project.  Construction support was also granted by
the United States National Science Foundation under Award PHY-0619727 and by the University of Rochester.  
Support for participating scientists was provided by NSF and DOE (USA), by CAPES and CNPq (Brazil), by CoNaCyT (Mexico), 
by Proyecto Basal FB 0821, CONICYT PIA ACT1413, Fondecyt 3170845 and 11130133 (Chile), by CONCYTEC, DGI-PUCP and IDI/IGI-UNI (Peru)
by the Latin American Center for Physics (CLAF) and by RAS and the Russian Ministry of Education and Science (Russia). 
We thank the MINOS Collaboration for use of its near detector data.  We acknowledge the dedicated work of the Fermilab 
staff responsible for the operation and maintenance of the beamline and detector, 
and we thank the Fermilab Computing Division for support of data processing.

This manuscript has been authored by Fermi Research Alliance, LLC under Contract No. DE-AC02-07CH11359 with the U.S. Department
of Energy, Office of Science, Office of High Energy Physics. The United States Government retains and the publisher, by
accepting the article for publication, acknowledges that the United States Government retains a non-exclusive, paid-up,
irrevocable, world-wide license to publish or reproduce the published form of this manuscript, or allow others to do so, for
United States Government purposes.


\end{document}